# Interface properties and built-in potential profile of a LaCrO$_3$/SrTiO$_3$ superlattice determined by standing-wave excited photoemission spectroscopy

S-C Lin,[1,2,*] C-T Kuo,[1,2] R. B. Comes,[3,4] J. E. Rault,[5] J.-P. Rueff,[5] S. Nemšák,[6,7] A. Taleb,[5] J. B. Kortright,[2] J. Meyer-Ilse,[2] E. Gullikson,[2] P. V. Sushko,[3] S. R. Spurgeon,[3,8] M. Gehlmann,[2,6] M. E. Bowden,[9] L. Plucinski,[6] S. A. Chambers,[3] and C. S. Fadley[1,2,†]

[1]Department of Physics, University of California Davis, Davis, California 95616, United States

[2]Materials Sciences Division, Lawrence Berkeley National Laboratory, Berkeley, California 94720, United States

[3]Physical and Computational Sciences Directorate, Pacific Northwest National Laboratory, Richland, Washington 99352, United States

[4]Department of Physics, Auburn University, Auburn, Alabama 36849, United States

[5]Synchrotron SOLEIL, L'Orme des Merisiers, Saint-Aubin-BP48, 91192 Gif-sur-Yvette, France

[6]Peter Grünberg Institut PGI-6, Research Center Jülich, 52425 Jülich, Germany

[7]Advanced Light Source, Lawrence Berkeley National Laboratory, Berkeley, California 94720, United States

[8]Energy and Environment Directorate, Pacific Northwest National Laboratory, Richland, Washington 99352, United States

[9]Environmental Molecular Sciences Laboratory, Pacific Northwest National Laboratory, Richland, Washington 99354, United States

**Corresponding Author**

*S.-C. Lin. E-mail: shclin@ucdavis.edu (S.-C.L.)

†C. S. Fadley. E-mail: fadley@physics.ucdavis.edu (C.S.F.)

**ABSTRACT:** LaCrO$_3$ (LCO) / SrTiO$_3$ (STO) heterojunctions are intriguing due to a polar discontinuity along [001], exhibiting two distinct and controllable charged interface structures [(LaO)$^+$/(TiO$_2$)$^0$ and (SrO)$^0$/(CrO$_2$)$^-$] with induced polarization, and a resulting depth-dependent potential. In this study, we have used soft- and hard-x-ray standing-wave excited photoemission spectroscopy (SW-XPS) to quantitatively determine the elemental depth profile, interface properties, and depth distribution of the polarization-induced built-in potentials. We observe an alternating charged interface configuration: a positively charged (LaO)$^+$/(TiO$_2$)$^0$ intermediate layer at the LCO$_{top}$/STO$_{bottom}$ interface and a negatively charged (SrO)$^0$/(CrO$_2$)$^-$ intermediate layer at the STO$_{top}$/LCO$_{bottom}$ interface. Using core-level SW data, we have determined the depth distribution of species, including through the interfaces, and these results are in excellent agreement with scanning transmission electron microscopy and electron energy loss spectroscopy (STEM-EELS) mapping of local structure and composition. SW-XPS also enabled deconvolution of the LCO and STO contributions to the valence band (VB) spectra. Using a two-step analytical approach involving first SW-induced core-level binding energy shifts and then valence-band modeling, the variation in potential across the complete superlattice is determined in detail. This potential is in excellent agreement with density-functional theory models, confirming that this method as a generally useful new tool for interface studies.

## I. INTRODUCTION

Interfaces between two distinct complex oxide materials offer a wide range of emergent electronic, magnetic, and optical properties that are not found in bulk materials. These include two-dimensional electron gases (2DEGs) in many coupled materials [1], interfacial ferromagnetism in materials that do not exhibit bulk ferromagnetism [2], and interface-induced photoconductivity due to interfacial dipole moments [3,4]. Superlattices (SLs) of these materials offer additional degrees of control and measurement because they are comprised of many repeating interfaces, thus amplifying interface-specific effects. For example, oxide SLs have produced the first observation of a polar vortex in $PbTiO_3/SrTiO_3$ (STO) SL [5], and a room-temperature multiferroic exhibiting ferroelectricity, and ferromagnetism in $LuFeO_3/LuFe_2O_4$ SLs [6]. It has recently been demonstrated by Comes *et al*. [7] that interfacial engineering can be used to induce a polarization in $LaCrO_3$ (LCO)/STO SLs. In light of recent studies of the $LaFeO_3$/STO(001) interface where promising photoconductive and photocatalytic behaviors have been observed [8,9] modulating the electronic structure and band alignment of a material in the form of a SL could be a promising avenue for light capture and conversion applications. To explain the behavior of these materials, an accurate experimental determination of the depth-dependent composition, electronic structure, and possible built-in potential gradients at buried interfaces in such SLs is essential. This paper demonstrates that standing-wave excited photoemission can uniquely and non-destructively determining the built-in potential, along with the other properties mentioned above.

While computational modeling at the level of density functional theory (DFT) enables predictions of electronic behavior in these materials, it is significantly more difficult to experimentally determine the depth profiles of composition, electronic structure and potential profiles in a SL. Traditional approaches for single-interface heterostructures cannot be readily applied to understand the behavior of systems consisting of multiple buried interfaces. In the case of a single interface, it is straightforward to measure electronic band alignment between a thin film and the underlying substrate using x-ray photoelectron spectroscopy (XPS) [10,11].

Careful modeling of the XPS data can even allow for determination of surface band bending and potential gradients due to band offsets across an interface [8,12]. When studying a SL, however, one obtains signals from multiple buried interfaces in a single measurement, making modeling exceptionally difficult due to the large number of assumptions that must be made to determine the properties of specific interfaces.

Standing-wave excited x-ray photoelectron (SW-XPS) measurements are a particularly promising way to overcome the challenges associated with SLs because they offer a mean to highlight individual interfaces by selectively tuning the intensity of the electric field with depth in the film [13,14]. This approach was first applied to an oxide SL by Gray *et al*. in particular for $La_{0.7}Sr_{0.3}MnO_3$/STO SL to study interfacial magnetic phenomena [15], and has since been used by Nemšák *et al*. to determine the depth distribution of the 2DEG in $GdTiO_3$/STO SLs [16].

In this work, we use SW-XPS to study the composition profile, band alignment, and built-in potential of an interface-engineered STO/LCO SL. We find that the electrostatic potential varies in both the STO and LCO layers of the SL indicating that there are distinct induced electric fields in the two oxides. A novel method of analyzing core-level shifts with SW excitation is used to derive the associated potential gradients in each layer.

## II. EXPERIMENT

### A. Sample Synthesis

The LCO/STO SLs were synthesized by oxide molecular beam epitaxy on conducting Nb-doped STO(001) substrates using a shuttered growth approach [7]. The Nb-doped STO(001) substrate was etched using boiling deionized water and annealed at 1000 °C for 30 minutes in an open-air tube furnace to produce a $TiO_2$-terminated surface. Prior to growth of the SL the flux of each element from the effusion cells was calibrated using a quartz crystal oscillator. Pure STO and LCO calibration films were then grown to more precisely adjust the flux of each element by monitoring the oscillations from RHEED during the shuttered growth [17]. After calibration the effusion cells were left hot and the substrates were heated to 600 °C in an ECR oxygen plasma to clean the surface of adventitious carbon. The film was then grown sequentially using one elemental source at a time to produce an SL structure consisting of [5 u.c. LCO/10 u.c. STO]$_{x10}$. By shuttering the individual metal beams, the SL was synthesized to have alternating positively-charged $(LaO)^+/(TiO_2)^0$ and negatively-charged $(SrO)^0/(CrO_2)^-$ interfaces, terminating with a $(CrO_2)^-$ layer at the free surface.

### B. Standing-Wave Excited Photoemission

In this method, the SW is created by the interference between incident and reflected x-rays, with the incidence angle $\theta_x$ being scanned over the first-order Bragg condition of the SL under study, as given by $\lambda_x = 2d_{ML}\sin\theta_B$. Here $\lambda_x$ is the wavelength of incident photon, $d_{ML}$ is the period of the SL and $\theta_B$ is the incidence angle for first-order Bragg reflection. The resulting SW electric field intensity varies sinusoidally with sample depth, with a period for first-order reflection that is very close to $d_{ML}$, which is 56.8 Å for our SL sample, the configuration of which is shown in Fig. 1(a). Scanning the incidence angle over the Bragg condition changes the position of the SW by half a cycle, and it is this variation that provides unique phase-sensitive depth resolution that is not possible with other modes of XPS. The vertical movement of the SW through the sample with changing incidence angle will thus enhance or reduce photoemission from different depths, generating what we will call a rocking curve (RC) of intensity that will have sensitivity to the depth distribution of individual elements, as illustrated below. Figure 1(a) shows a schematic view of such a SW measurement for our specific sample configuration, with different parameters and angles defined. A final important point is that the amplitude of the SW modulation is proportional to the square root of the reflectivity (R). It is thus useful to maximize R by, for example, tuning the photon energy to be near a strong absorption resonance for one of the elements within the sample [15,16]. Finally, a specially-written x-ray optics computer code (Yang x-ray Optics, YXRO) is used in analyzing our SW-XPS data [13,14].

SW-XPS measurements were performed at beamline Cassiopee of SOLEIL synchrotron, with the angle $\theta_{xe} = 45°$, as defined in Fig. 1(a), and hard x-ray SW-XPS measurements were performed at beamline Galaxies of SOLEIL synchrotron, with an angle of $\theta_{xe} = 90°$. The radiation polarization was in the photoemission plane in both cases. The energy resolution of the soft x-ray SW-XPS is 500 meV and that of hard x-ray SW-XPS is 440 meV. X-ray absorption measurements were carried out at Cassiopee using total yield and at beamline 6.3.2 of the Advanced Light Source by direct reflectivity.

### C. Scanning transmission electron microscopy and electron energy loss spectroscopy measurements

Samples were prepared for STEM-EELS using a FEI Helios NanoLab Dual-Beam Focused Ion Beam (FIB) microscope and a standard lift out procedure, with initial cuts made at 30 kV and final polishing done at 5 kV / 5.5° and 2 kV / 6° incidence angle. STEM-HAADF images and STEM-EELS maps were collected along the STO [100] zone-axis on an aberration-corrected JEOL ARM-200CF microscope operating at 200 kV, with a convergence angle of 27.5 mrad and an EELS collection angle of 82.7 mrad. Spectra were collected with a 1 Å spot size, 1 eV ch$^{-1}$ energy dispersion, and a 4x energy binning to improve the signal collection rate. No plural scattering correction was performed since zero loss measurements confirm that the samples are sufficiently thin (t/λ ≈ 0.5 IMFP). The composition maps were processed using principal component analysis (PCA) to further reduce noise.

## III. RESULTS AND DISCUSSION

### A. Standing-Wave Excited Photoemission and Rocking Curves

We conducted two sets of soft x-ray experiments with photon energies just below and just above the La $M_5$ x-ray absorption maximum at 830.5 eV, as shown in Fig. 1(b). As illustrated in Fig. 1(c), the real (refractive) and imaginary (absorptive) parts of the index of refraction, delta and beta, respectively, of the LCO layer vary dramatically in the proximity of the absorption peak. Two photon energies, 829.7 eV and 831.5 eV, were chosen to maximize reflectivity at two positions adjacent to the absorption peak, as discussed in more detail in the Supplementary Information (Fig. S1). Most importantly, this choice of photon energies results in a shift in the SW phase between two measurements, as illustrated in Figs. 1(d) and 1(e), and enlarges the range of sampling depth for the SW-XPS experiments to encompass more or less the first bilayer of the sample. Figures 1(f) and 1(g) also demonstrate more clearly the true sampling depth, with the SW intensities being multiplied by the appropriate inelastic mean free paths (IMFPs) for the representative photoelectron peaks (note the logarithmic scale).

In order to shift the SW along the depth direction, spectra were measured as a function of incidence angle between 5.5° and 10° for hv = 829.7 eV, between 6° and 10° at 831.5 eV. The first-order Bragg reflection from the multilayer is spanned in all cases. To illustrate the spatial distribution of SW versus incidence angle, the YXRO-derived electric field intensities as a function of incidence angle and sample depth are shown in Figs. 1(d) and 1(e) for the photon energies of 829.7 eV and 831.5 eV. In Fig. 1(d), at 829.7 eV, as the incidence angle increases, in the angle range of 5.5° to 7°, the maximum of the SW lies near the first interface, which we designate as LCO$_{top}$/STO$_{bottom}$. The maximum then sweeps down to the middle of the first STO layer in the angle range of 7° to 8° and stays there until the end of the angle scan. On the other hand, the movement of the SW in Fig. 1(e) at 831.5 eV shows similar behavior as in Fig 1(d) but with an overall downward shift of ~20 Å, yielding more sensitivity to the second interface, $STO_{top}/LCO_{bottom}$. Note that the simulated electric field intensities are all normalized to the incident beam intensity.

Combining SW results from Figs 1(d) and 1(e), and the estimated depth sensing in Figs. 1(f) and 1(g) that allow for inelastic scattering, we see that in light of the short IMFPs of the valence electrons excited with soft x-rays (~18 Å for STO layer and ~16 Å for LCO layer), SW-XPS yields strong sensitivity to the top LCO layer and first interface ($LCO_{top}/STO_{bottom}$). In order to probe more deeply, we have also taken a complementary set of hard x-ray SW data at an energy of 3.5 keV. For this case, the angle scan over the Bragg region is between 1.2° and 2.6°. The mean IMFPs of our hard x-ray data are 50 Å, and roughly equal to $d_{ML}$= 56.8 Å. This means ~90% of the photoemission yields are from the top two SL periods, so our data at this energy samples the first two buried interfaces. The corresponding simulation-derived electric field strength distribution and photoemission yield at this higher energy are also shown in Figure S2.

To first determine the detailed depth-resolved composition of the sample, we have measured the RCs of the most intense core levels for each atomic species in the LCO/STO SL at photon energies of 829.7 eV, 831.5 eV and 3.5 keV. Figure 2(a) shows the strongest core-level spectra for all atomic species in the LCO/STO SL and their fitted components at hν = 829.7 eV. Here we see C 1s, O 1s, La 4d, Cr 3p, Sr 3d and Ti 2p spectra, with their soft x-ray RCs as derived from peak-fitted intensities shown in Fig 2(b). The effects of the resonant La excitation are seen in the La 4d and Sr 3d spectra. There are strongly screened final states (green) for the La $4d_{5/2}$ and $4d_{3/2}$ manifolds that are shifted ~3.3 eV to higher binding energy from the unscreened doublet (blue) [18]. We have used the sum of these two doublets to obtain the RC in Fig. 2(b). Also, a prominent high-binding-energy shoulder in the Sr 3d spectrum is a $4d^{-1}5p^{-1}4f$ resonant Auger peak associated with La [19]; its intensity was subtracted in arriving at the Sr 3d RC. In contrast, the spectra of Cr 3p and Ti 2p are relatively simple. The low- and high- binding energy peaks in Cr 3p result from well-known multiplet splittings involving both magnetic and spin-orbit interactions [12]. Significantly, in the Ti 2p spectrum, there is only a $Ti^{4+}$ component and no evidence of a lower-binding-energy $Ti^{3+}$ shoulder. In addition to the dominant O 1s peak (green) corresponding to oxygen in the SL, a surface-related component (magenta) is present, most likely due to surface OH formation resulting from the exposure to atmosphere in transferring the sample to the measurement chamber [20]; its RC is in fact found to be very similar to that of C 1s, another surface-associated species, so we do not plot it in Figs. 2(b) and 2(c).

Figures 2(b) and 2(c) present the experimental RCs (open circles) and best-fit simulations from our x-ray optical program [13] (curves) of the representative elemental states at photon energies of 829.7 eV and 831.5 eV. For the C 1s, La 4d, Cr 3p and Ti 2p spectra in Fig. 2(a), because the blue and green components share the same spatial distribution, the sums of their intensities are plotted as the RCs. In contrast, only the green components are taken into account for O 1s and Sr 3d. A linear background is subtracted from the experimental RCs to compensate the intensity variation of the incident photon resulted from slightly off-axis sample rotation. Note that all the RCs are normalized to a maximum of unity and are offset vertically for readability. The fractional modulation of each RC can thus be read directly from the ordinate scale.

In Figs. 2(b) and 2(c), the RCs of the core levels for the atomic species in the same layer, e.g. La 4d and Cr 3p, as well as Sr 3d and Ti 2p, have almost identical intensity profiles; conversely the RCs corresponding to different layers are completely out of phase, e.g. La 4d and Sr 3d. At the same time, the C 1s RCs exhibit unique profiles owing to its unique location at the surface.

The RCs of O 1s follow those of La 4d and Cr 3p since most of the photoemission yield of O 1s comes from the topmost LCO layer when measuring with soft x-ray excitation. The same conclusions are reached by looking at the deeper-probing RCs with 3.5 keV excitation in Fig. 2(d), although the O1s RC tends to be rather flat, since averaging over RCs in a few bilayers. Note the generally excellent agreement between experiment and simulation for the RCs at all energies, in which the thicknesses of all layers and the degree of interfacial mixing have been varied over a number of choices to yield the best fit as judged by R-factor, with a number of prior SW photoemission studies suggesting an accuracy of ~ ±2-3 Å [13,21].

It is noteworthy that the shapes of the two soft x-ray RCs change markedly in going from below (Fig. 2(b)) to above (Fig. 2(c)) the La 3d resonance; thus, the two sets of data are fully complementary. We also find very strong modulations in these soft x-ray experimental RCs of up to 70 %, which facilitates measuring and fitting experiment to theory accurately, including the small phase differences between the different RCs, thus finally arriving at the optimal SL structure determination. For example, we find that there are very small phase differences of 0.2° between Sr 3d and Ti 2p RCs and 0.1° between La 4d and Cr 3p RCs at $h\nu$ = 831.5 eV, suggesting asymmetric atom distributions among the two constitute elements of the STO and LCO layers. The effect is smaller, but still noticeable, at $h\nu$ = 829.7 eV, with reduced magnitude due to its different probing profile, as discussed above. The conclusion of asymmetric interfacial structures, e.g. between the top and bottom of STO, is consistent with the previous STEM study reported by Comes *et al*. [22].

As noted above, we show in Fig. 2(d) SW-XPS measurements obtained at 3.5 keV. These data probe more deeply and yield information on the top two interfaces as discussed above. Here, we again see excellent agreement between experiment and simulation, and for exactly the same sample structure that we determined with the softer x-ray energies. Moreover, Bragg peaks along with Kiessig fringes are clearly seen in the hard x-ray data. The relative positions and amplitudes of Kiessig fringes with respect to the Bragg peak are very sensitive to thickness gradients in the SL [15,16]. Hence, the agreement between experiment and simulation ensures excellent regularity for the whole SL. The corresponding simulation-derived electric field strength distribution and photoemission yield maps at 3.5 keV are shown in Figure S2.

The simulated RCs have been calculated using the YXRO program [13], with appropriate x-ray optical parameters, IMFPs, and various trial sample structures as input. The SL structure was optimized by minimizing the error between all experimental and simulated RCs simultaneously via iteratively adjusting the input SL structure. The SL structures resulting from the best-fit simulations of the soft x-ray data, Figs. 2(a) and 2(b), and the hard x-ray data, Fig. 2(c), are found to be the same. Figure 3 shows the optimized SL structure as determined by SW-XPS and compares this structure to that from STEM-EELS maps, which have been obtained from the same sample. In the SW-XPS structure (Fig. 3(a)), we find that there is a 9Å thick surface contamination layer (C+O) at the surface. Moreover, from the SW-XPS results, we find around ~2-3-Å-thick interfaces in this SL, which consist of alternating positively and negatively charged structures: $(LaO)^+/(TiO_2)^0$ with positive charge (green) at the $LCO_{top}/STO_{bottom}$ interface, and $(SrO)^0/(CrO_2)^-$ with negative charge (yellow) at the $STO_{top}/LCO_{bottom}$ interface. This result is consistent with an A cation layer/B cation layer stacking sequence at both kind of interfaces. The spatial distributions of Sr, Ti, Cr and La determined by SW-XPS are plotted separately in Fig. 3(b), using the same color scheme as in the STEM-EELS maps in Fig. 3(c). In Fig. 3(b), a spatial offset between the distributions of A and B cations is clearly resolved; the spatial distributions of La and Sr atomic species are offset ~2 Å from those of Cr and Ti. These results can be directly compared to the STEM-EELS composition map, where agreement regarding the asymmetric nature of the two interfaces is seen. A grayscale high-angle annular dark field (HAADF) STEM image is shown along with the STEM-EELS composition maps in Fig. 3(c). These images demonstrate an overall excellent quality and regularity of the SL and reveal no apparent structural imperfection. Moreover, from Fig. 3(a), we notice that the thickness of the SW-XPS derived LCO plus half of the charged interfaces is ~ 18Å. This is about 8% lower than the 19.4Å expected, based on the bulk LCO lattice constant. However, judging from the STEM-EELS and HAADF images, 5 complete u.c. of LCO are clearly resolved in most of the repeat units and no atomic planes is obviously missing. Therefore, the thickness variation relative to bulk would likely propagate to step edges and have a negligible effect on the physics that we are going to exam in the following. Further information regarding the structure and uniformity of sample, including integrated profiles of STEM-EELS composition maps, HAADF images with various magnification and reflectivity measurements, can found in Figures S3, S4 & S6.

## B. SW Derived Depth-resolved Built-in Potential

With a SL structure with alternating positively and negatively charged interfaces, one might ask does the resulting parallel-plate-capacitor-like interfacial configuration lead to electric fields across the interfaces and through the layers? If so, how do these fields modify the electronic structure along the interface normal, in particular the valence-band maximum (VBM)? To answer these questions, we have simultaneously measured the valence-band spectra and the core-level peak positions as the incidence angle is varied. Combining these two data sets permits a unique determination of the layer-dependent densities of states, as well as the depth-resolved potential. These results are summarized in Figures 4 and 5.

We first show in Figs. 4(a) and 4(b) the valence band RCs at photon energies of 829.7 eV and 831.5 eV, which clearly exhibit much different SW behavior as the angle is increased. Then, expanding upon prior work by our group [23] and other group using harder x-rays at few keV [24,25] by simultaneously analyzing the valence-band (VB) and layer-specific core-level RCs, the VB contributions from the LCO or STO components of the SL can be distinguished. Since we are probing with soft x-rays, nearly all the intensities detected in Figs. 4(a) and 4(b) are emitted from the topmost LCO/STO interface, with the LCO contributing the majority. There are three prominent features in the VB spectra of Fig. 4(a) at low angles that we label A, B, and C. Based on prior DFT calculations, these correspond to the bonding states of the Cr 3d spin-up $t_{2g}$ band, the nonbonding O 2p states and bonding states of Cr 3d and O 2p, respectively [26,27]. Moreover, we assume that VB spectra are the sum of matrix-element-weighted DOSs (MEWDOSs) for all constituent layers, attenuated by the photoelectron IMFPs. Noting that the intensities at each binding energy step in the VB spectra contain contributions from both the LCO and STO layers, a given RC can be represented as a linear combination of RCs from the individual layers [23], and can be written as:

$$I_{VB}(E_b, \theta_x) = \sum_{\text{layer } j} \rho_j(E_b) \times I_j(\theta_x) \qquad (1)$$

Here $I_{VB}(E_b, \theta_x)$ is the experimental RC intensity at a binding energy $E_b$ and x-ray incidence angle $\theta_x$, $j = $ LCO or STO, $I_j(\theta_x)$ is the SW RC contribution from a layer $j$, for which we use Cr 3p for LCO and Ti 2p for STO, and $\rho_j(E_b)$ are the deconvolution coefficients related directly to the MEWDOS in layer $j$. The valence-band RCs at each energy step have been fitted to a linear combination of the characteristic RCs by a least-square fitting routine. Finally, the layer-projected MEWDOSs are derived via weighting the angular integrated valence-band spectra of Figs. 4(a) and 4(b) with the coefficients derived by fitting over the whole binding energy range.

Figure 4(c) shows the angle-integrated valence-band spectra and the corresponding projected MEWDOSs for the different constituent layers at photon energies of 829.7 eV and 831.5 eV. The valence-band edges for the projected MEWDOSs are determined by linear extrapolation to zero, as shown schematically by the black dashed lines in Fig. 4(c). Figure 4(d) shows for reference the MEWDOS results from conventional XPS measurements for thick-film LCO and bulk STO (single crystal substrate). Furthermore, an interface-induced state, as annotated as peak D in the STO MEWDOSs, which is not seen in the bulk STO electronic structure is revealed by the deconvolution. This it is due to a combination of Cr diffusion into STO [28,29], and possibly a slight artifact of the deconvolution procedure. We define the maximum of state E as the valence-band edge of the projected STO MEWDOS in order to directly compare it to the valence-band spectra of bulk STO in the following discussion. When the photon energy is switched from 829.7 eV to 831.5 eV, we find that the projected MEWDOSs of LCO and STO both shift toward lower binding energy: the valence-band edges move from 0.9 eV to 0.7 eV and 3.3 eV to 3.0 eV for LCO and STO, respectively. The fact that the energy levels of the MEWDOS of both constituent layers vary with changes in the SW-XPS depth profile unambiguously reveals that variations in the electrostatic potential are present within both LCO and STO.

We now discuss a novel method for determining the detailed form of the built-in potential as a function of depth, beginning with analysis of the variation of core-level binding energies as the SW is scanned through the SL. Figures 5(a) and 5(b) show the experimental peak shifts for the major components in the Sr 3d and La 4d core-level spectra versus incidence angle at photon energies of 829.7 eV and 831.5 eV, along with simulated results. The components used for analyzing the experimental peak shifts are the Sr $3d_{3/2}$ feature and the screened feature in the La $4d_{5/2}$ spectrum, with their positions determined by curve fitting (Figure S6). The experimental variations for the Sr 3d and La 4d peaks have small, but reproducible changes in binding energy of the order of 0.1-0.2 eV as the incidence angle is scanned. Moreover, the form of these is quite different for the two x-ray energies, as expected from the different phases and forms of the SW. Note that we focus on the *change* in potential rather than its absolute value for now, and we represent the peak positions by their energy separation relative to the average peak position over the angle scan.

We have modeled the spectra of these peaks over the entire incidence angle range and then extracted the angular dependence of their maximum position as the simulated peak shift. Here we assume that the core-level binding energy follows this potential at each depth, tracking perfectly with the VB maximum in that layer, as in the method of Kraut *et al.* [11], and further that the potential can be described as a linear variation within each layer. Using the accurate depth-dependent photoemission intensity from Figs. 1(f) and 1(g), we have simulated the peak shifts in the La 4d and Sr 3d spectra, representing core levels in LCO and STO. The intensity versus binding energy in a given layer $j$ at depth $z_i$ with an incidence angle $\theta_x$, $I_j(E_b, \theta_x, z_i)$, where $j$ denotes LCO or STO and $i$ a continuous depth variable within each layer, is described for convenience as a Voigt function with FWHM equal to the estimated experimental energy resolution, $V(E_b - E_{b,j}^{lin}(z_i))$. Here $E_{b,j}^{lin}(z_i)$ is the linear built-in potential shift of the binding energy at a given depth in layer $j$. The photoemission intensity from depth $z_i$ is the product of the field strength and the inelastic attenuation factor, $|E(z_i, \theta_x)|^2 \exp(-z_i / \Lambda_e \sin\theta_e)$, with $\theta_x$ being the incidence angle, $\Lambda_e$ the IMFP, and $\theta_e$ the electron exit angle with respect to the surface, given by

$\theta_e = \theta_x + 45°$. Thus, the binding energy variation as a function of x-ray incidence angle $\theta_x$, $I_{j,\max}(E_b, \theta_x)$, is calculated from the maximum intensity position of the sum, and is described as,

$$I_{j,\max}(E_b, \theta_x) = \text{maximum of } \sum_{z_i} I_{j,\max}(E_b, \theta_x, z_i) = \sum_{z_i} V(E_b - E_{b,j}^{lin}(z_i)) |E(z_i, \theta_x)|^2 \exp(-z_i / \Lambda_e \sin\theta_e) \quad . \quad (2)$$

Then using the accurate depth-dependent photoemission intensity from Figs. 1(f) and 1(g) as the second two factors in the RHS of this equation, as well as the assumed linear form of the potential (the first factor) as a trial-and-error input, the best potential gradients were determined by least-square fitting, and these result in the smooth curves shown in Figs. 5(a) and 5(b). More details on this simulation method are contained in the discussion of Figures S7 and S8.

We find generally excellent agreement between experiment and theory in Figs. 5(a) and 5(b), with only Sr 3d showing less variation in theory than in experiment, perhaps due to intermixing with the LCO layer. The potential gradients yielding these fits are shown in Figs. 5(e) and 5(f) and include +1eV and a -0.8 eV changes in binding energy along the depth direction within the LCO and STO layers, respectively.

The energy steps or valence-band offsets at each interface shown in Fig. 5(e) are further determined by the following analysis of the valence-band maxima. Figures 5(c) and 5(d) summarize two different ways of looking at the overall VB spectra at the same two photon energies. The deconvoluted MEWDOSs of the STO layer and LCO layer from Fig 4(c) is one set of curves. The curves denoted "simulation" are based upon inserting the XPS bulk reference spectra from Fig. 4(d), $I_{VB,j}^{XPS}(E_B)$, with $j$ = LCO or STO, into a sum over the built-in potential similar to that shown in Equation (2),

$$I_{VB,j}(E_b) = \sum_{\theta_x} \sum_{z_i} I_{VB,j}^{XPS}(E_b - E_b^0(z_i)) |E(z_i, \theta_x)|^2 \exp(-z_i / \Lambda_e \sin\theta_e) , \quad (3)$$

with the total potential $E_b^0(z_i)$ shown in Fig. 5(e), including potential gradients within constituent layers and steps at the polar interfaces due to band offsets, with the steps being varied to fit the VBM shifts discussed above. A further elaboration of this simulation process can be found in Figure S9 and its discussion.

By combining the derivation of the slopes of electrostatic potential within each layer and the magnitude of valence band offsets at two kinds of charged interface, we finally determine the absolute potential value with respect to the VB maxima, annotated as the SW-XPS derived profile in Figs. 5(e) and 5(f). We note that this procedure yields a uniquely precise specification of the potential variations along the depth direction. The VB edge of the LCO layer shifts toward higher binding energy by 1 eV within 5 u.c. of LCO, which results in a change in binding energy from 0.2 eV at the $STO_{top}$/$LCO_{bottom}$ (negatively charged) interface, or the surface for the topmost LCO, to 1.2 eV at the $LCO_{top}$/$STO_{bottom}$ (positively charged) interface. At the same time, the VB edge of the STO layer shifts to a lower binding energy by 0.8 eV within 10 u.c. STO, which is equivalent to a change in binding energy from 3.1 eV at the $LCO_{top}$/$STO_{bottom}$ interface (positively charged) to 2.3 eV at the $STO_{top}$/$LCO_{bottom}$ interface (negatively charged). This result indicates clear agreement between the qualitative expectation of the charged-interface configuration and the signs of the potential gradients: higher (lower) binding energy for valence electrons at the positively (negatively) charged interfaces.

### C. Density Functional Theory

We have corroborated these results using DFT simulations with the PBEsol density functional [30], as implemented in the VASP code [31,32] with an adjustable $U_{eff}$ parameter for d-d correlation in both layers and these results are found to agree excellently with the experimental results as to both slopes and offsets at the interfaces, as shown by the black curves in Figs 5(e) and 5(f). In Fig. 5(e), the $U_{eff}$ values in LCO (8 eV) and STO (3 eV) were chosen to yield the correct bulk bandgaps. We note that while $U_{eff}$ (Cr) = 3.0 eV and $U_{eff}$ (Ti) = 8.0 eV produces a correct trend and that theory agrees with experiment to within about 0.5 eV within the layers, the best agreement between the calculated and the experimental VB maximum profiles is found for $U_{eff}$ (Cr) = 1.5 eV and $U_{eff}$ (Ti) = 4.0 eV, as shown in Fig. 5(f). This may indicate that the larger values of $U_{eff}$ introduce artificial electronic structure effects that exaggerate the internal field, or that the interfaces contain defects that partially offset the correlation effects on the field in the film. To see the trend of how the VB maximum profiles vary with the values of $U_{eff}$, a further discussion on these theoretical calculations with different choices can be found in Figure S10.

## IV. CONCLUSION

In summary, standing-wave excited soft- and hard- x-ray photoemission measurements have been applied to a $LaCrO_3/SrTiO_3$ SL that is expected to contain charged interfaces, in order to extract the depth-resolved atomic and electronic structure, and for the first time, the built-in potential. In the soft x-ray measurements, two photon energies above and below the La $M_5$ absorption edge were carefully chosen. These values lead to very large reflectivities and thus RC modulations of up to 70% and, because of the different phases of the SW with depth at the two energies, a sampling range which covers nearly the entire top LCO/STO bilayer, including top and bottom interfaces. In addition, complementary hard x-ray measurements were conducted to increase the probing depth. In all of these experiments, the Bragg peak is clearly resolved in the RCs, and for the higher energy x-ray, also Kiessig fringes. The same depth distributions for each atomic species are derived from RC analysis of the soft and hard x-ray regimes, and these distributions are in excellent agreement with STEM-EELS composition maps. Both sets of RC data, along with the STEM-EELS maps, are consistent with alternating charged interfaces: a

$(LaO)^+/(TiO_2)^0$ intermediate layer at the $LCO_{top}/STO_{bottom}$ interface and a $(SrO)^0/(CrO_2)^-$ intermediate layer at the $STO_{top}/LCO_{bottom}$ interface. Furthermore, we have deconvoluted the valence-band spectra into the MEWDOS of STO and LCO layers by analyzing the layer-specific, core-level RCs together with valence-band RCs. Further sequential analysis of core-level shifts as the SW is scanned vertically with angle, and the deconvoluted VB spectra compared to reference simulations, has permitted determining in unique detail the variation of the built-in potential with depth, including the band offsets at the polar interfaces. This overall potential is in excellent agreement with DFT theory, confirming the method. As a final comment, we believe that the SW methods we have introduced here should have wide applicability in the study of not only oxide interfaces and their built-in potentials, but also many other types of heterostructures, including e.g. the electrochemical double layer, for which similar core-level shifts with SW excitation have been observed recently, but not yet analyzed with the method introduced here [33].

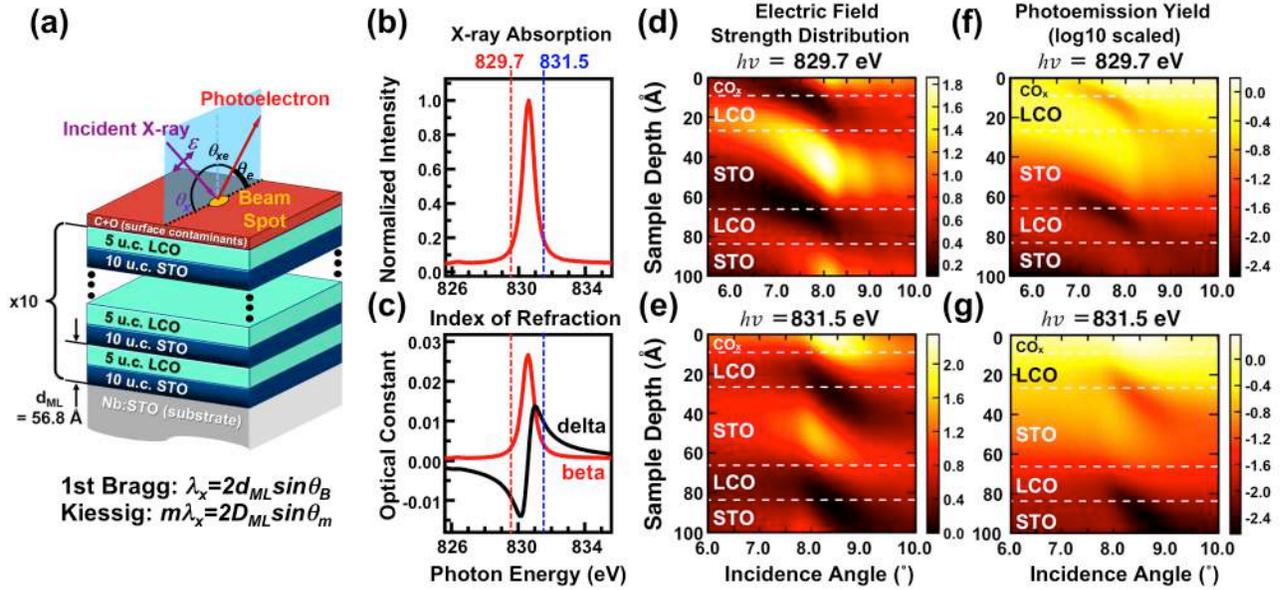

FIG. 1. (a) Schematic of the superlattice made up of 10 bilayers of LCO and STO, consisting of 5 unit cells of LCO, 17.6 Å thick, and 10 unit cells of STO, 39.2 Å thick, grown epitaxially on a Nb-doped STO(001) substrate. The two sources of standing-wave structure in the rocking curves are indicated: Bragg reflection from the multilayer with period $d_{ML}$ and Kiessig fringes associated with the full thickness of the multilayer stack $D_{ML}$. (b) The x-ray absorption coefficient over the La $M_5$ edge. (c) The real (delta) and imaginary (beta) parts of the index of refraction, as derived by Kramers-Kronig analysis. To enhance the reflectivity and thus the strength of the standing wave effect, two photon energies were chosen, below and above the La $M_5$ absorption maximum. The electric field strength distribution derived from x-ray optics calculations at these two energies, (d) 829.7 eV and (e) 831.5 eV as a function of sample depth and incidence angle. Note the significant shift in position between the two energies. The corresponding calculated photoemission yields with depth, (f) and (g), plotted on log10 scales.

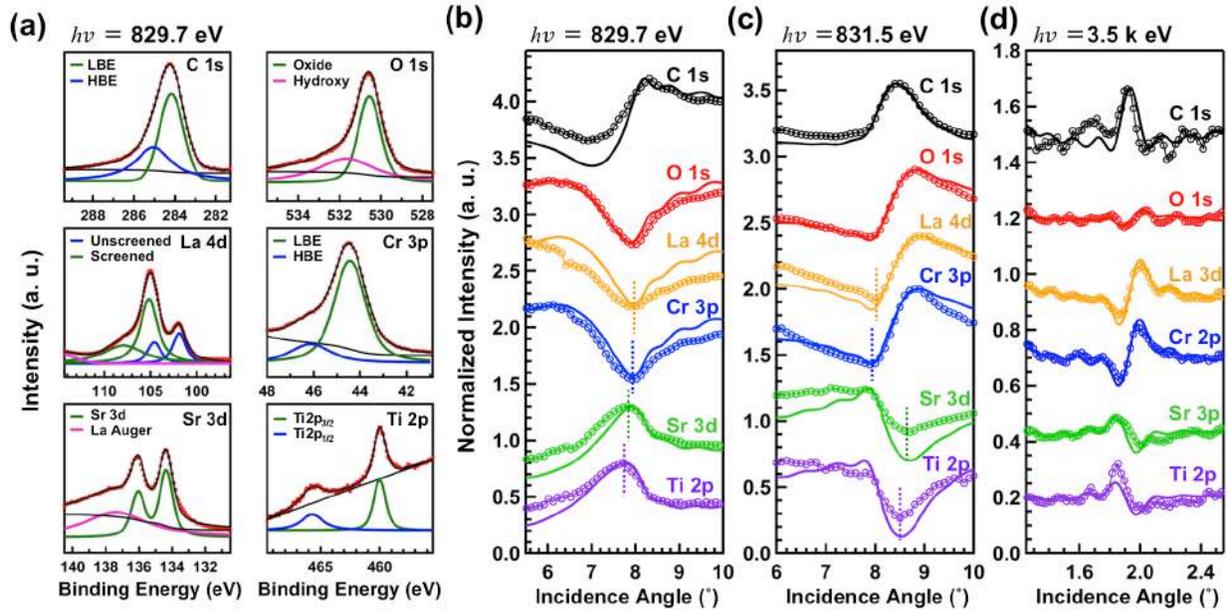

FIG. 2. (a) Experimental spectra and fitted components of the strongest core level for each atomic species in the LCO/STO superlattice at a photon energy of 829.7 eV. In several cases in (b) and (c), the intensities used are the sums of blue and green components in (a). Experimental (open circles) and YXRO simulated (solid) rocking curves of representative elemental states at photon energies of (b) 829.7 eV, (c) 831.5 eV. The dashed vertical lines indicate the phase difference of the rocking curves. (d) As (b) and (c) but for experimental and simulated rocking curves at a photon energy of 3.5 keV. Note that in the case of 3.5 keV, clear Bragg peaks and Kiessig fringes are visible in both experiment and theory.

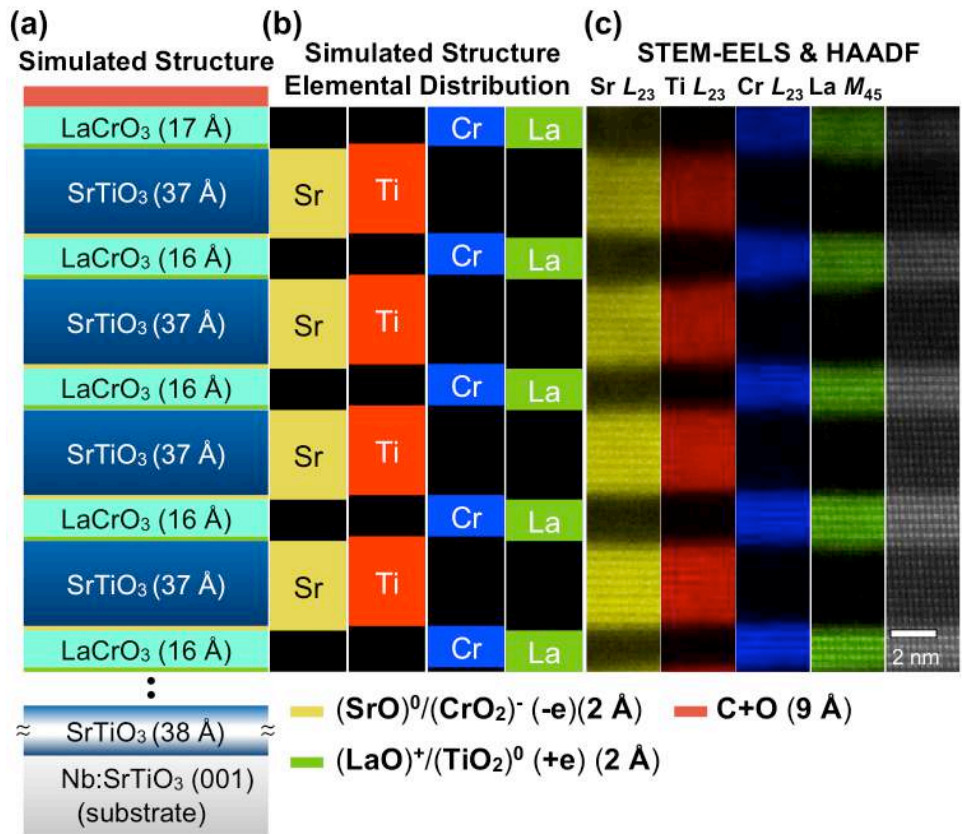

FIG. 3. (a) The sample structure determined via fitting YXRO simulations of both the hard and soft x-ray SW rocking curves in Figs. 2(b), (c) and 3(d). The models used for interface interdiffusion at the $LCO_{top}/STO_{bot}$ and $STO_{bot}/LCO_{top}$ interfaces and surface contamination layer are indicated below the main panel. (b) The separate depth profiles of major atomic species in the LCO/STO superlattice derived from YXRO. (c) Corresponding principal component analysis-filtered STEM-EELS composition maps and a representative STEM-HAADF image of the LCO/STO superlattice. The color codes of EEELS are yellow, red, blue and green for the Sr $L_{23}$, Ti $L_{23}$, Cr $L_{23}$ and La $M_{45}$ absorption edges, respectively. The HAADF is shown in greyscale.

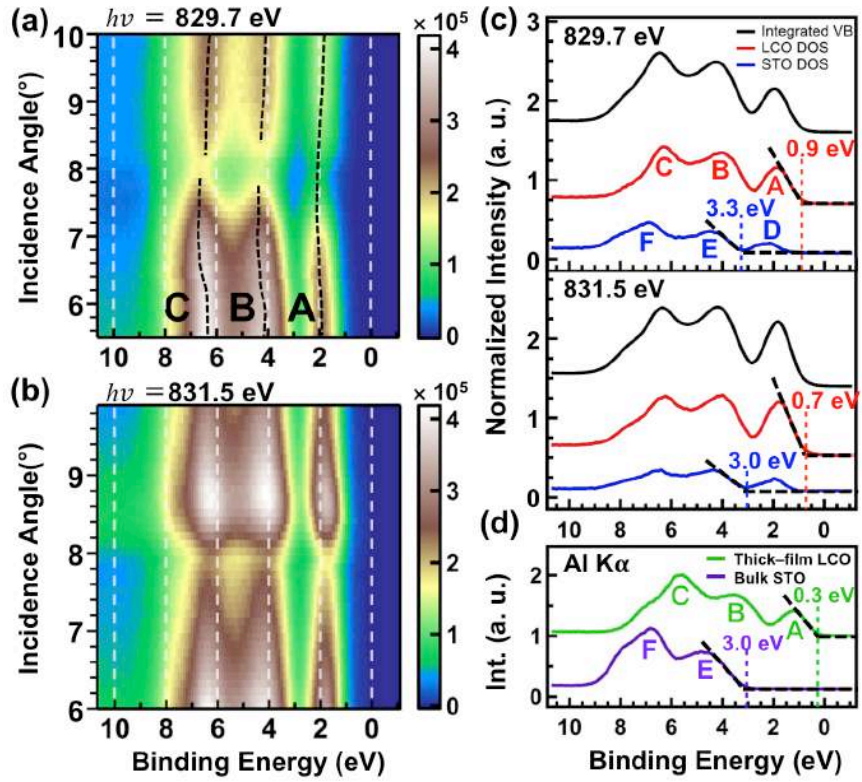

FIG. 4. Experimental RCs for the superlattice valence-band spectra at photon energies of (a) 829.7 eV and (b) 831.5 eV. (c) Angle integrated spectra for (a) and (b) (black curves) and corresponding decomposed LCO-like (red curves) and STO-like (blue curves), representing matrix-element-weighted densities of states (MEWDOSs). (d) Reference XPS valence band spectra of bulk STO (single crystal substrate) and thick-films LCO acquired with Al Kα (1486.6 eV).

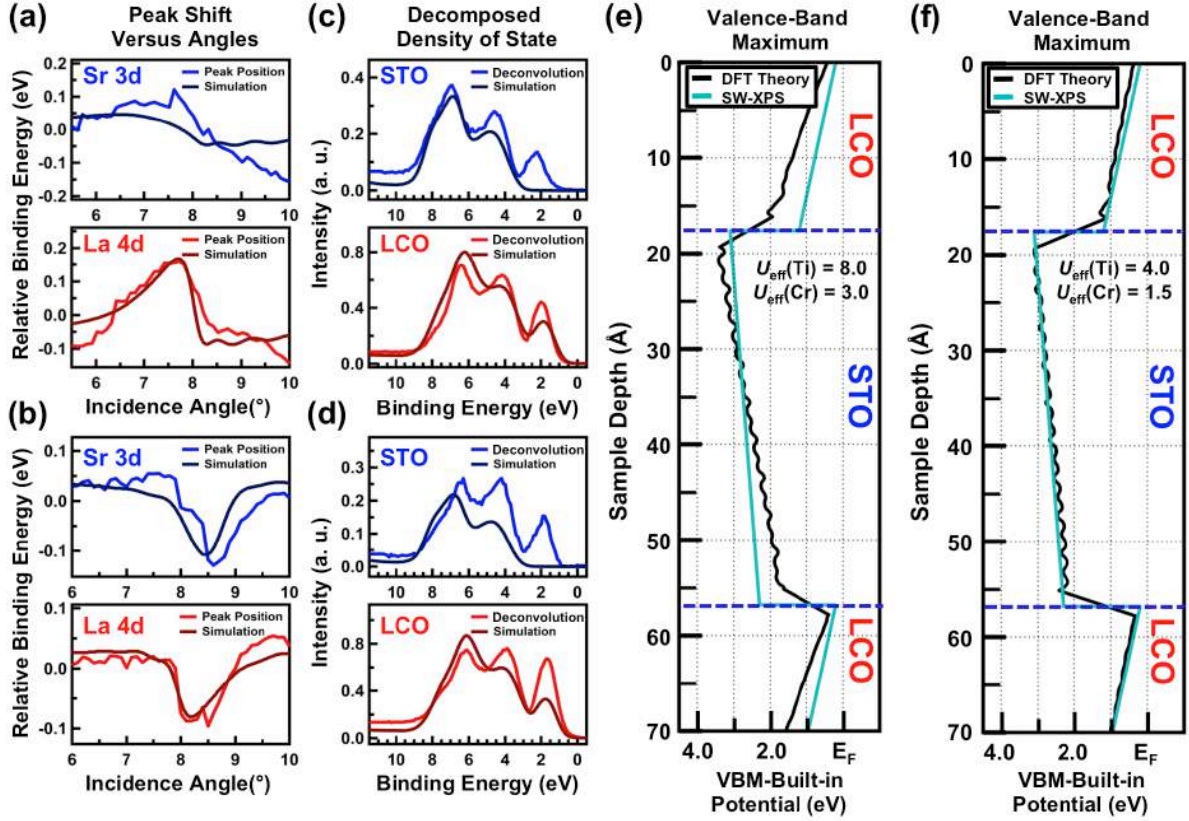

FIG. 5. Experimental and simulated relative peak shifts for Sr 3d and La 4d core levels versus incidence angle at photon energies of (a) 829.7 eV and (b) 831.5 eV. Experimental valence-band decompositions, showing the contributions from the STO and LCO layers, and corresponding simulations using XPS reference spectra from bulk STO and thick-film LCO, at photon energies of (c) 829.7 eV and (d) 831.5 eV. (e),(f) SW-XPS derived (turquoise curves) and DFT calculated (PBEsol) depth-resolved valence band maximum (black curves) for the top three layers of the LCO/STO superlattice. This SW-XPS derived depth profile is determined by optimizing the simulations in (a)-(d). The DFT theoretical profile is calculated in (e) with $U_{eff}$(Ti) 8.0 eV and $U_{eff}$(Cr) = 3.0 eV to match the bulk bandgaps in STO and LCO, and in (f) with $U_{eff}$(Ti) 4.0 eV and $U_{eff}$(Cr) = 1.5 eV, which yields the best fit to experiment.

# ACKNOWLEDGEMENT

This work was supported by the US Department of Energy under Contract No. DE-AC02-05CH11231 (Advanced Light Source), and by DOE Contract No. DE-SC0014697 through the University of California Davis (salary for C.-T.K, S.-C.L. and C.S.F.). C.S.F. has also been supported for salary by the Director, Office of Science, Office of Basic Energy Sciences (BSE), Materials Sciences and Engineering (MSE) Division, of the U.S. Department of Energy under Contract No. DE-AC02-05CH11231, through the Laboratory Directed Research and Development Program of Lawrence Berkeley National Laboratory, through a DOE BES MSE grant at the University of California Davis from the X-Ray Scattering Program under Contract DE-SC0014697, through the APTCOM Project, "Laboratoire d'Excellence Physics Atom Light Matter" (LabEx PALM) overseen by the French National Research Agency (ANR) as part of the "Investissements d'Avenir" program, and from the Jűlich Research Center, Peter Grűnberg Institute, PGI-6, through a joint Jülich/LBNL collaboration. The work at PNNL was supported by the U.S. Department of Energy, Office of Science, Division of Materials Sciences and Engineering under Award #10122 as well as the PNNL Linus C. Pauling Distinguished Postdoctoral Program, and was performed in the Environmental Molecular Sciences Laboratory, a national scientific user facility sponsored by the Department of Energy's Office of Biological and Environmental Research and located at PNNL.

# REFERENCES

[1] S. Stemmer and S. J. Allen, Two-dimensional electron gases at complex oxide interfaces, Annu. Rev. Mater. Res. **44**, 151–171 (2014).

[2] J. Hoffman, I. C. Tung, B. B. Nelson-Cheeseman, M. Liu, J. W. Freeland, and A. Bhattacharya, Charge transfer and interfacial magnetism in $(LaNiO_3)n/(LaMnO_3)_2$ Superlattices, Phys. Rev. B **88**, 144411 (2013).

[3] K. Nakamura, H. Mashiko, K. Yoshimatsu, and A. Ohtomo, Impact of built-in potential across $LaFeO_3/SrTiO_3$ heterojunctions on photocatalytic activity, Appl. Phys. Lett. **108**, 211605 (2016).

[4] M. Nakamura, F. Kagawa, T. Tanigaki, H. S. Park, T. Matsuda, D. Shindo, Y. Tokura, and M. Kawasaki, Spontaneous polarization and bulk photovoltaic effect driven by polar discontinuity in $LaFeO_3/SrTiO_3$ heterojunctions, Phys. Rev. Lett. **116**, 156801 (2016).

[5] A. K. Yadav, C. T. Nelson, S. L. Hsu, Z. Hong, J. D. Clarkson, C. M. Schlepüetz, A. R. Damodaran, P. Shafer, E. Arenholz, L. R. Dedon, *et al.*, Observation of polar vortices in oxide superlattices, Nature **530**, 198–201 (2016).

[6] J. A. Mundy, C. M. Brooks, M. E. Holtz, J. A. Moyer, H. Das, A. F. Rébola, J. T. Heron, J. D. Clarkson, S. M. Disseler, Z. Liu, A. Farhan, *et al.*, Atomically engineered ferroic layers yield a room-temperature magnetoelectric Multiferroic, Nature **537**, 523 (2016).

[7] R. B. Comes, S. R. Spurgeon, S. M. Heald, D. M. Kepaptsoglou, L. Jones, P. V. Ong, M. E. Bowden, Q. M. Ramasse, P. V. Sushko, and S. A. Chambers, Interface-induced polarization in $SrTiO_3$-$LaCrO_3$ superlattices, Adv. Mater. Interfaces **3**, 201500779 (2016).

[8] R. B. Comes and S. A. Chambers, Interface structure, band alignment, and built-in potentials at $LaFeO_3/n–SrTiO_3$ heterojunctions, Phys. Rev. Lett. **117**, 226802 (2016).

[9] S. R. Spurgeon, P. V. Sushko, R. B. Comes, and S. A. Chambers, Dynamic interface rearrangement in $LaFeO_3/n–SrTiO_3$ heterojunctions, Phys. Rev. Mater. **1**, 63401 (2017).

[10] R. B. Comes, P. Xu, B. Jalan, and S. A. Chambers, Band alignment of epitaxial $SrTiO_3$ thin films with $(LaAlO_3)_{0.3}$-$(Sr_2AlTaO_6)_{0.7}$ (001), Appl. Phys. Lett. **107**, 131601 (2015).

[11] E. A. Kraut, R. W. Grant, J. R. Waldrop, and S. P. Kowalczyk, Precise determination of the valence-band edge in x-ray photoemission spectra: application to measurement of semiconductor interface potentials, Phys. Rev. Lett. **44**, 1620 (1980).

[12] S. A. Chambers, L. Qiao, T. C. Droubay, T. C. Kaspar, B. W. Arey, and P. V. Sushko, Band alignment, built-In potential, and the absence of conductivity at the $LaCrO_3/SrTiO_3$(001) heterojunction, Phys. Rev. Lett. **107**, 206802 (2011).

[13] S.-H. Yang, A. X. Gray, A. M. Kaiser, B. S. Mun, B. C. Sell, J. B. Kortright, and C. S. Fadley, Making use of x-ray optical effects in photoelectron-, Auger electron-, and x-ray emission spectroscopies: total reflection, standing-wave excitation, and resonant effects, J. Appl. Phys. **113**, 073513 (2013).

[14] C. S. Fadley, S. Nemšák, Some future perspectives in soft- and hard- X-ray photoemission, J. Electron Spectrosc. Relat. Phenom. **195**, 409 (2014).

[15] A. X. Gray, C. Papp, B. Balke, S.-H. Yang, M. Huijben, E. Rotenberg, A. Bostwick, S. Ueda, Y. Yamashita, K. Kobayashi, *et al.*, Interface properties of magnetic tunnel junction $La_{0.7}Sr_{0.3}MnO_3/SrTiO_3$ superlattices studied by standing-wave excited photoemission spectroscopy, Phys. Rev. B **82**, 205116 (2010).

[16] S. Nemšák, G. Conti, A. X. Gray, G. K. Palsson, C. Conlon, D. Eiteneer, A. Keqi, A. Rattanachata, A. Y. Saw, A. Bostwick, *et al.*, Energetic, spatial, and momentum character of the electronic structure at a buried interface: The two-dimensional electron gas between two metal oxides, Phys. Rev. B **93**, 245103 (2016).

[17] J. H. Haeni, C. D. Theis, and D. G. Schlom, RHEED intensity oscillations for the stoichiometric growth of $SrTiO_3$ thin films by reactive molecular beam epitaxy, J. Electroceram. **4**, 385–391 (2004).

[18] W.-Y. Howng, and R. Thorn, The multicomponent structure of the 4d orbital in X-ray photoelectron spectra of the lanthanum(III) ion, J. Chem. Phys. Lett. **15**, 463 (1978).

[19] P. Lagarde, A.-M. Flank, H. Ogasawara, and A. Kotani, Resonant photoemission of La and Yb at the 3d absorption edge, J. Electron Spectrosc. Relat. Phenom. **128**, 193 (2003).

[20] P. Scheiderer, F. Pfaff, J. Gabel, M. Kamp, M. Sing, and R. Claessen, Surface-interface coupling in an oxide heterostructure: Impact of adsorbates on $LaAlO_3/SrTiO_3$, Phys. Rev. B **92**, 195422 (2015).

[21] S.-H. Yang, K. Balke, C. Papp, S. Döring, U. Berges, L. Plucinski, C. Westphal, C. M. Schneider, S. S. P. Parkin, and C. S. Fadley, Determination of layer-resolved composition, magnetization, and electronic structure of an Fe/MgO tunnel junction by standing-wave core and valence photoemission, Phys. Rev. B **84**, 184410 (2011).

[22] R. B. Comes, S. R. Spurgeon, D. M. Kepaptsoglou, M. H. Engelhard, D. E. Perea, T. C. Kaspar, Q. M. Ramasse, P. V. Sushko, and S. A. Chambers, Probing the origin of interfacial carriers in $SrTiO_3$–$LaCrO_3$ superlattices, Chem. Mater. **29**, 1147–1155 (2017).

[23] A. M. Kaiser, A. X. Gray, G. Conti, J. Son, A. Greer, A. Perona, A. Rattanachata, A. Y. Saw, A. Bostwick, S. Yang, *et al.*, Suppression of near-Fermi level electronic states at the interface in a $LaNiO_3$/$SrTiO_3$ Superlattice, Phys. Rev. Lett. **107**, 116402 (2011).

[24] J. C. Woicik, E. J. Nelson, and P. Pianetta, Direct measurement of valence-charge asymmetry by x-ray standing waves, Phys. Rev. Lett. **84**, 773 (2000).

[25] J. C. Woicik, E. J. Nelson, D. Heskett, J. Warner, L. E. Berman, B. A. Karlin, I. A. Vartanyants, M. Z. Hasan, T. Kendelewicz, Z. X. Shen, and P. Pianetta, X-ray standing-wave investigations of valence electronic structure, Phys. Rev. B **64**, 125115 (2001).

[26] K. P. Ong, P. Blaha, and P. Wu, Origin of the light green color and electronic ground state of $LaCrO_3$, Phys. Rev. B **77**, 073102 (2008).

[27] P. V. Sushko, Q. Liang, M. Bowden, T. Varga, G. J. Exarhos, F. K. Urban III, D. Barton, and S. A. Chambers, Multiband optical absorption controlled by lattice strain in thin-film $LaCrO_3$, Phys. Rev. Lett. **110**, 077401 (2013).

[28] S. A. Chambers, M. Gu, P. V. Sushko, H. Yang, C. Wang, and N. D. Browning, Ultralow contact resistance at an epitaxial metal/oxide heterojunction through interstitial site doping, Adv. Mater. **25**, 4001–4005 (2013).

[29] R. B. Comes, P. V. Sushko, S. M. Heald, R. J. Colby, M. E. Bowden, and S. A. Chambers, Band-gap reduction and dopant interaction in epitaxial La,Cr Co-doped $SrTiO_3$ Thin Films, Chem. Mater. **26**, 7073–7082 (2014).

[30] J. P. Perdew, A. Ruzsinszky, G. I. Csonka, O. A. Vydrov, G. E. Scuseria, L. A. Constantin, X. L. Zhou, and K. Burke, Restoring the density-gradient expansion for exchange in solids and surfaces, Phys. Rev. Lett. **100**, 136406 (2008).

[31] G. Kresse and J. Furthmüller, Efficient iterative schemes for ab initio total-energy calculations using a plane-wave basis set, Phys. Rev. B **54**, 11169 (1996).

[32] G. Kresse and J. Hafner, Ab initio molecular-dynamics simulation of the liquid-metal–amorphous-semiconductor transition in germanium, Phys. Rev. B **49**, 14251 (1994).

[33] S. Nemšák, A. Shavorskiy, O. Karslioglu, I. Zegkinoglou, P. K. Greene, E. C. Burks, K. Liu, A. Rattanachata, C. S. Conlon, A. Keqi, *et al*. Concentration and chemical-state profiles at heterogeneous interfaces with sub-nm accuracy from standing-wave ambient-pressure photoemission, Nat. Commun. **5**, 5441 (2014).

# Supporting Information

# Interface properties and built-in potential profile of a LaCrO$_3$/SrTiO$_3$ superlattice determined by standing-wave excited photoemission spectroscopy

S-C Lin,[1,2,*] C-T Kuo,[1,2] R. B. Comes,[3,4] J. E. Rault,[5] J.-P. Rueff,[5] S. Nemšák,[6,7] A. Taleb,[5] J. B. Kortright,[2] J. Meyer-Ilse,[2] E. Gullikson,[2] P. V. Sushko,[3] S. R. Spurgeon,[3,8] M. Gehlmann,[2,6] M. E. Bowden,[9] L. Plucinski,[6] S. A. Chambers,[3] and C. S. Fadley[1,2,†]

[1]Department of Physics, University of California Davis, Davis, California 95616, United States

[2]Materials Sciences Division, Lawrence Berkeley National Laboratory, Berkeley, California 94720, United States

[3]Physical and Computational Sciences Directorate, Pacific Northwest National Laboratory, Richland, Washington 99352, United States

[4]Department of Physics, Auburn University, Auburn, Alabama 36849, United States

[5]Synchrotron SOLEIL, L'Orme des Merisiers, Saint-Aubin-BP48, 91192 Gif-sur-Yvette, France

[6]Peter Grünberg Institut PGI-6, Research Center Jülich, 52425 Jülich, Germany

[7]Advanced Light Source, Lawrence Berkeley National Laboratory, Berkeley, California 94720, United States

[8]Energy and Environment Directorate, Pacific Northwest National Laboratory, Richland, Washington 99352, United States

[9]Environmental Molecular Sciences Laboratory, Pacific Northwest National Laboratory, Richland, Washington 99354, United States

**Tuning photon energy to maximize reflectivity- Figure S1**

In Figure S1, we demonstrate how tuning photon energy over an absorption maximum, as done previously in SW-XPS [1,2], can enhance reflectivity, and thus standing-wave modulation. Fig. S1(a) shows a set of calculated reflectivities, based on the optical constants in Fig. 1(c) of the main text, with two clear maxima as the photon energy goes over the La $M_5$ resonance. Fig. S1(b) then shows the reflectivity maximum as a function of photon energy, with the two photon energies chosen for this experiment indicated, together with the absorption maximum. The energy of 831.5 eV was chosen slightly off the maximum to increase the vertical phase difference in the two standing-wave scans.

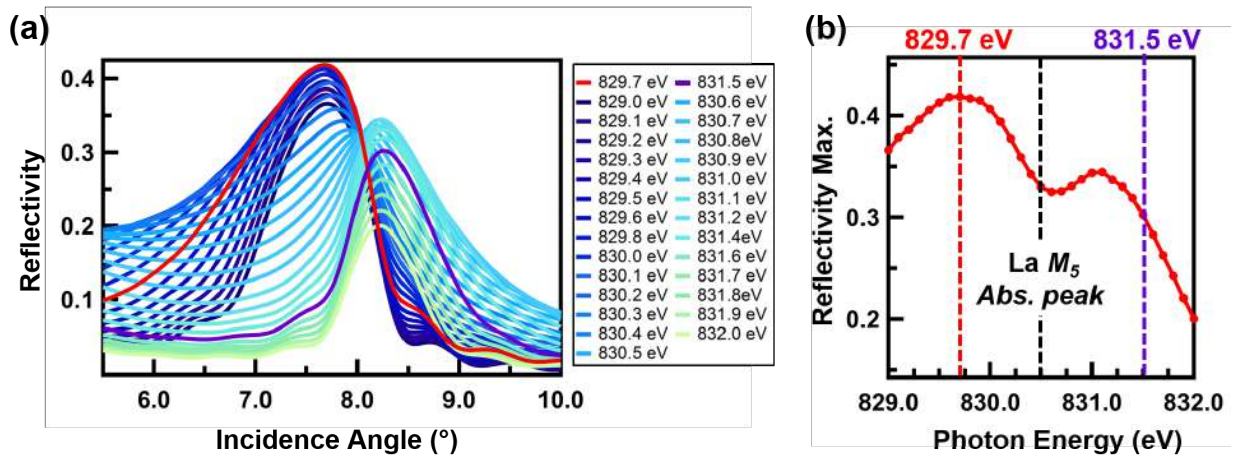

FIG. S1. (a) Simulated reflectivity as a function of incidence angle at various photon energies in the proximity of the La $M_5$ absorption. The reflectivity curves of the two experimental photon energies of the SW-XPS, 829.7 eV and 831.5 eV, are red and violet color coded. (b) The maximum of the reflectivity curves plotted as a function of photon energy, with our two photon energies indicated, together with the position of the La $M_5$ absorption peak.

**Hard x-ray photoemission results- Figure S2**

In Figs. S2(a) and S2(b) we show the calculated x-ray wave field and photoemission intensity profiles for 3.5 keV excitation, equivalent to those presented in Figs. 1(d) and 1(f) for 829.7 eV and Figs. 1(e) and 1(g) for 831.5 eV. Note that 3.5 keV is not a resonant energy, so the x-ray optical constants can be derived from online tabulations [3]. In Fig. S2(c), we show the core-level spectra, with representative peak fitting that was used to derive rocking curves (RCs). These can be compared to Fig. 2(a), noting that these high-energy spectra are generally simpler in form, with lower inelastic backgrounds, a simpler spin-orbit doublet for La 4d that involves deeper probing in the sample and less influence of either final-state screening or surface/interface effects. These spectra were analyzed to yield the RCs in Fig. 2(d).

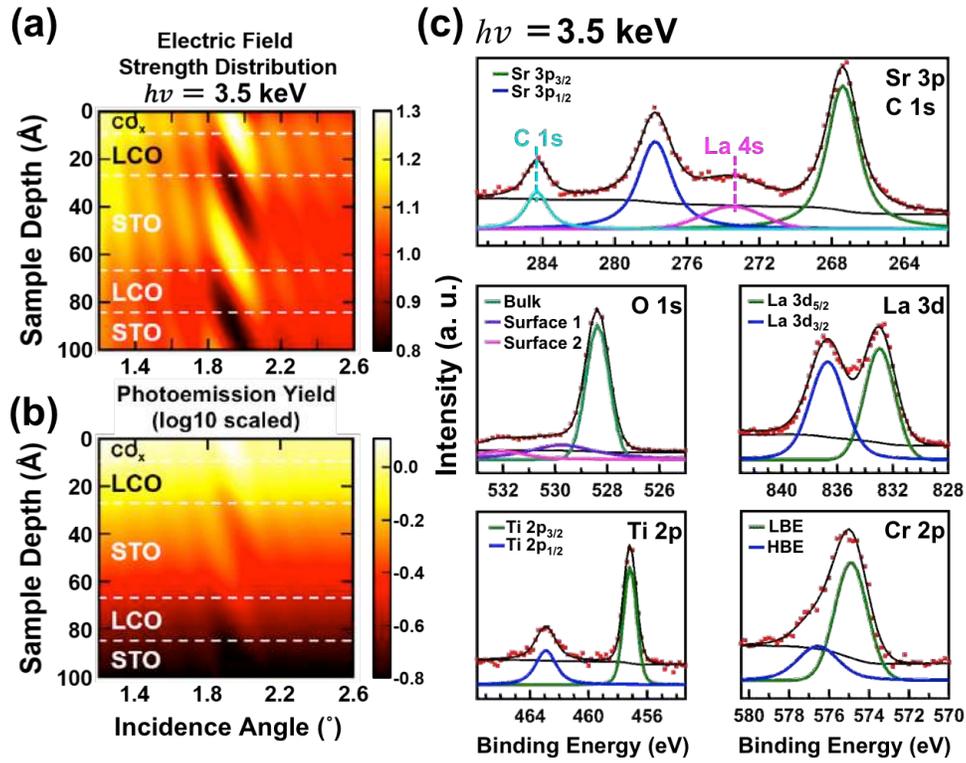

FIG. S2. (a) Simulated depth-resolved electric field strength distribution as a function of incidence angle at photon energy of 3.5 keV. (b) The corresponding photoemission intensity estimate plotted on a log10 scale. (c) Representative core level spectra with their fitted components, as used to derive the RCs in Fig. 2(d).

**STEM-EELS and HAADF images-Figures S3 & Figure S4:**

The high-angle annular dark field (HAADF) images acquired along with STEM-EELS measurements are shown at two different magnifications in Figure S3. These images along with the STEM-EELS mapping in the main text confirm an excellent quality and uniformity of the superlattice and reveal no apparent structural defects or imperfections.

Figure S4 shows representative STEM-EELS composition maps and integrated line profiles for the La $M_{45}$, Cr $L_{23}$, Sr $L_{23}$, and Ti $L_{23}$ edges. These integrated profiles indicate an intermixing of Sr atoms inside LCO layer and a finite Cr concentration within the STO layers. We note that the gradient in composition is the result of a wedge-like sample shape; furthermore, the increase in effective Sr signal at the sample surface (shown in the line profile), is the result of an overlap with the Pt $M_{45}$ edge present in the protective Pt capping layer, and so is not a genuine increase in Sr concentration.

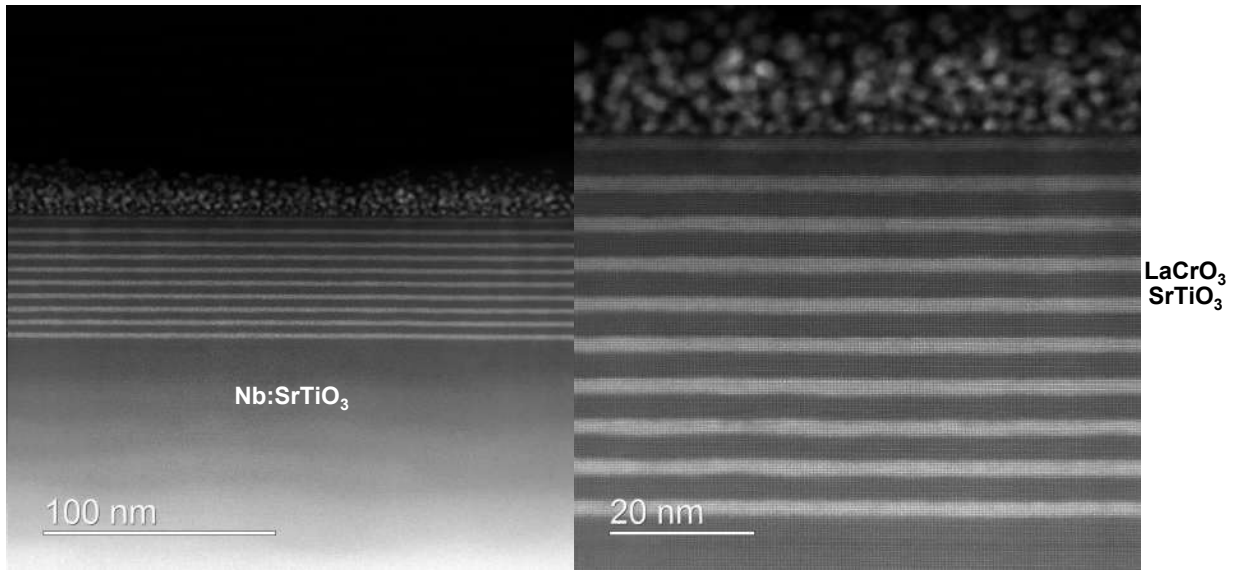

FIG. S3. Representative STEM-HAADF images taken along the STO [100] zone-axis, illustrating the excellent quality and uniformity of the superlattice.

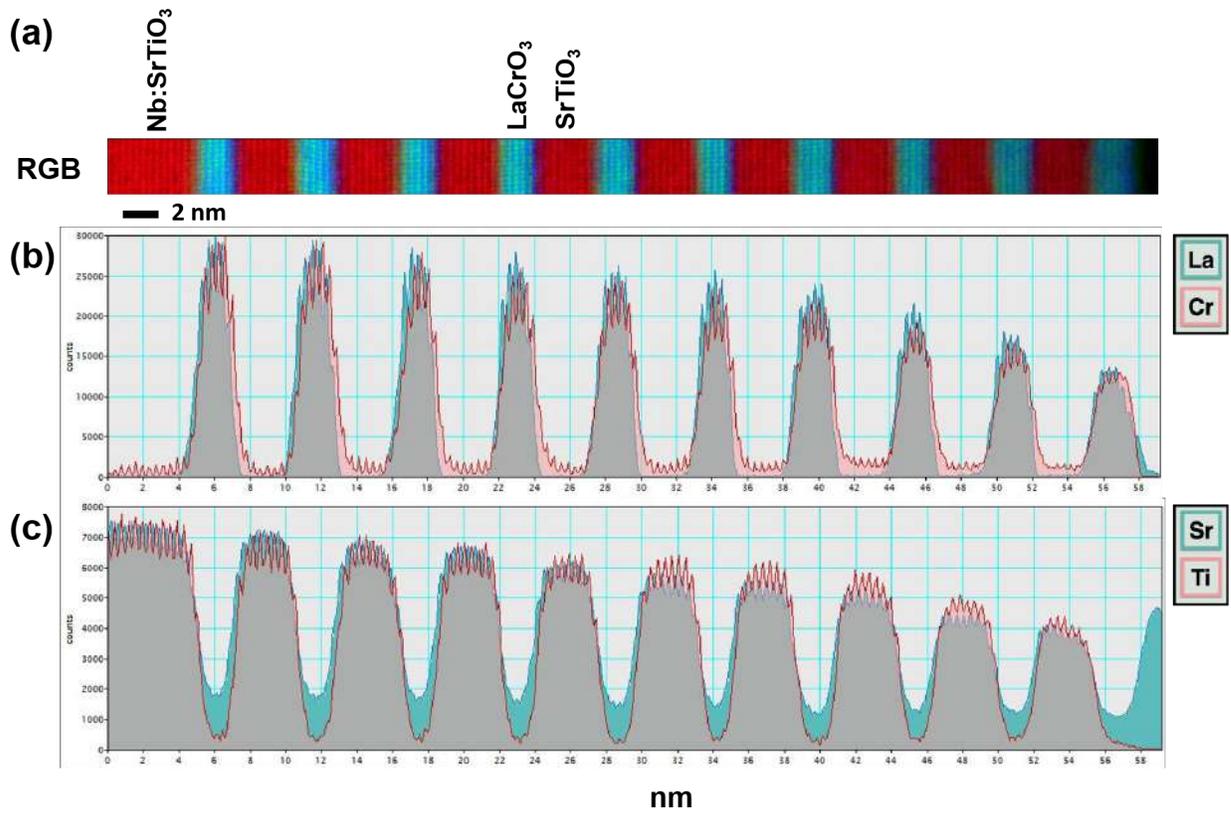

FIG. S4. (a) Representative STEM-EELS composition map and integrated line profiles for the (b) La $M_{45}$ and Cr $L_{23}$ (c) Sr $L_{23}$ and Ti $L_{23}$ edges.

**The concentration of Ti$^{3+}$ and its impact on the built-in potential - Figure S5:**

Since Ti$^{3+}$ is one of the major sources of mobile carrier of the widely studied oxide LAO/STO hetero-interface, an important question to ask is "what is the concentration of Ti$^{3+}$ in the LCO/STO SL and how it impact the built-in potential?"

In our study, the concentration of Ti$^{3+}$ state can be derived from two aspects of our measurements: the intensity of the Ti$^{3+}$ state in Ti 2p core level spectra and the extend of in-diffused Ti in the LCO layer. First, as shown in Fig. 2(a), of the main text, there no evident Ti$^{3+}$ shoulder observed in the Ti 2p spectrum. In the other hand, judging from the Ti L-edge EELS map below, Figure S5, the concentration of the in-diffused Ti in the LCO layer is ~1 at %, which is close to the uncertainty and detection limit of the EELS measurement. Both pieces of evidence indicate limited presence of Ti$^{3+}$ state in the LCO/STO SL. Moreover, according to a recent study by Comes et al.[4], it turns out that the built-in potential is preserved even in the presence of Ti$_{Cr}$ substitutional defects. Hence, we conclude that the existence of Ti$^{3+}$, if there is some, has no apparent impact on the built-in potential in LCO/STO SL.

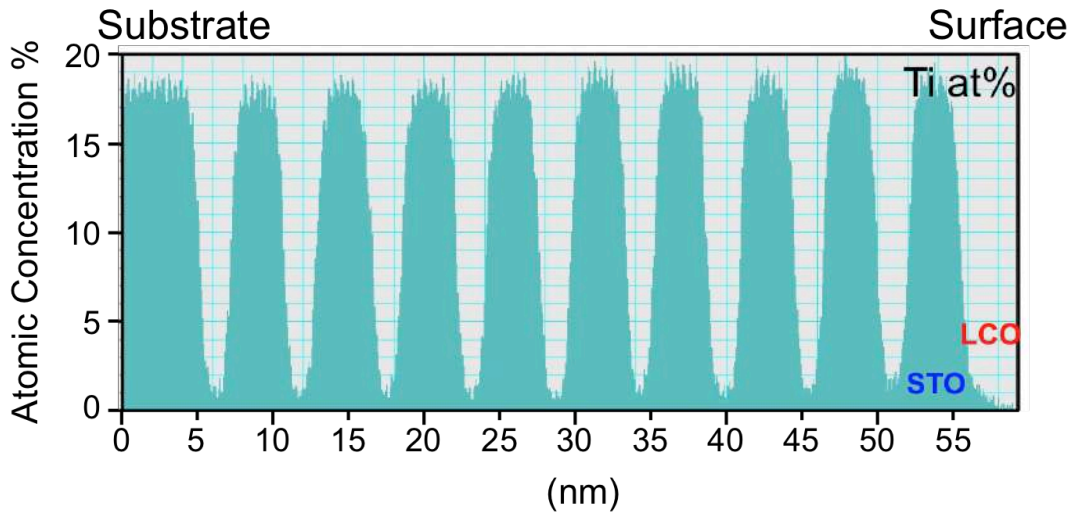

FIG. S5. Atomic concentration of Ti versus sample depth

**X-ray reflectivity measurements – Figure S6**

Lab based Cu Kα x-ray reflectivity (XRR) measurements were performed on the superlattice to estimate standing wave intensity and verify that the superlattice was grown as intended. The results of this measurement are shown in Figure S6, in comparison to modeling of structure that has a good agreement with the SW-XPS structure, confirm that the repeating structure of the superlattice is uniform throughout the thickness of the film, in good agreement with the STEM images shown above.

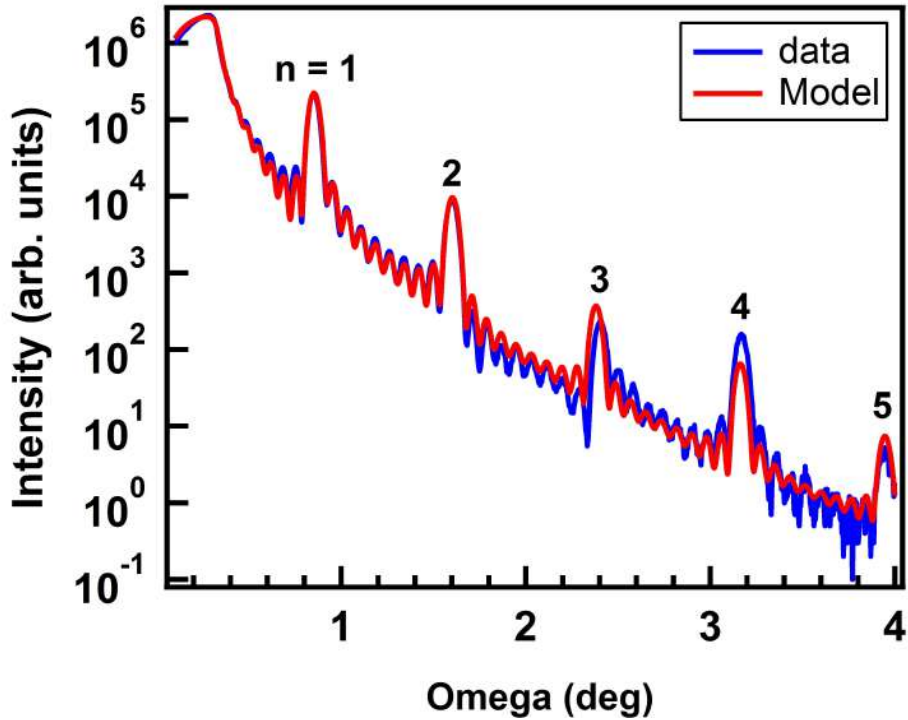

FIG. S6. Lab-based X-ray reflectivity (XRR) data and corresponding model calculations for the $STO_{10}/LCO_5$ superlattice.

**Determination of core-level rocking curves-- Figure S7:**

The spectra and fitting used to derive the energy shifts of the Sr $3d_{3/2}$ and La $4d_{5/2}$ (screened) peaks that have been analyzed and simulated in Figs. 5(a) and 5(b).

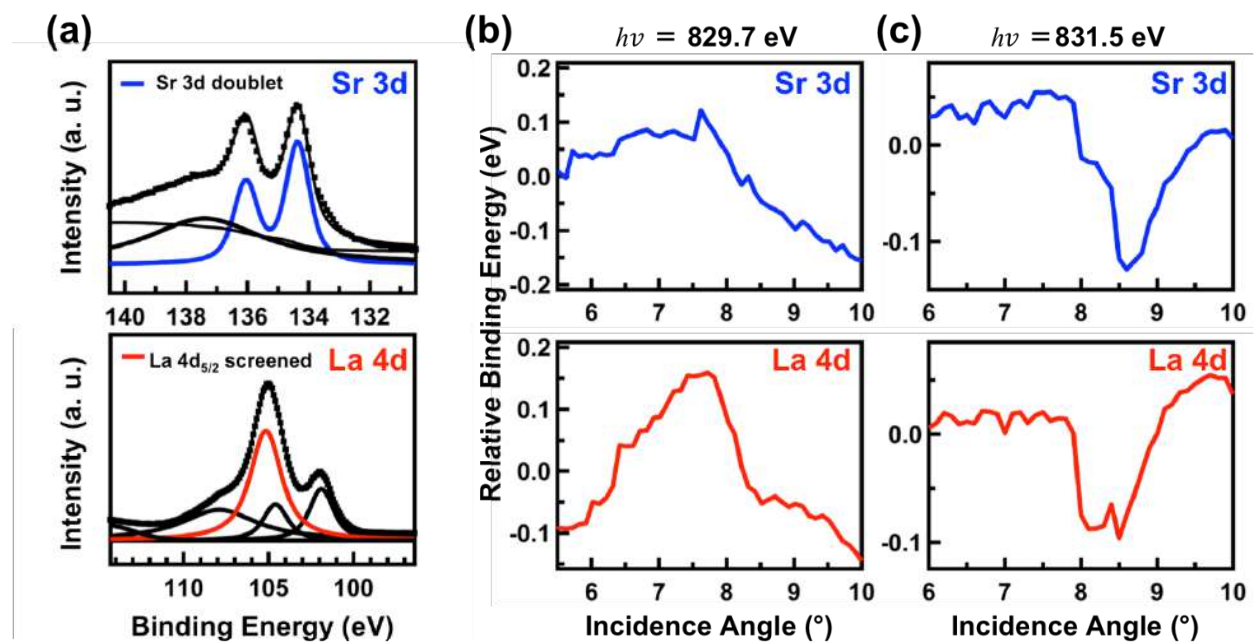

FIG. S7. (a) Experimental spectra and fitted components of La 4d and Sr 3d at a soft x-ray photon energy of 829.7 eV. (b) The relative binding energies of peaks in (a), Sr 3d doublet (blue) and screened La $4d_{5/2}$ (red), are plotted versus incidence angle at photon energies of (b) 829.7 eV and (c) 831.5 eV. Note the marked differences in shape.

**Determination of the built-in potential from core-level binding-energy shifts-- Figure S8:**

The detailed mathematical representation for simulating the peak shift of a core level in layer $j$ versus incidence angle, $I_{j,\max}(E_b, \theta_x)$, can be written in the form of Eq. (2) in the text as,

$$I_{j,\max}(E_b, \theta_x) = \text{maximum of } \sum_{z_i} I_{j,\max}(E_b, \theta_x, z_i) = \sum_{z_i} V(E_b - E_{b,j}^{lin}(z_i))|E(z_i, \theta_x)|^2 \exp(-z_i / \Lambda_e \sin\theta_e)$$

where $I_j(E_b, \theta_x, z_i)$ is the intensity versus binding energy of the simulated core level spectrum, in a given layer $j$ at depth $z_i$ with an incidence angle $\theta_x$, $j$ denotes LCO or STO and $i$ is a continuous depth variable within each layer. As shown in Fig. S8(a), a Voigt function, $V(E_b - E_{b,j}^{lin}(z_i))$, is used for convenience as a basis to simulate the core level spectrum, with the FWHM of the Voigt function taken from the experimental core level fitting. As shown in Fig. S8(b), $E_{b,j}^{lin}(z_i)$ is the built-in potential shift of the core level spectrum at a given depth in layer $j$, which is the trial input for optimizing the simulation. Meanwhile, the photoemission intensity from depth $z_i$ is the product of the field strength and the inelastic attenuation factor, $|E(z_i, \theta_x)|^2 \exp(-z_i / \Lambda_e \sin\theta_e)$, with $\theta_x$ being the incidence angle, $\Lambda_e$ the IMFP, and $\theta_e$ the electron exit angle (cf. Fig. 1(a) in the text). The simulated core level spectrum in layer $j$ at depth $z_i$, $I_j(E_b, \theta_x, z_i)$, is then derived via the sum in the above equation. In Fig. S8(c), the intensities $I_j(E_b, \theta_x, z_i)$ at four sample depths Z1 to Z4 are shown. In Fig. S8(d), via summing $I_j(E_b, \theta_x, z_i)$ over the all the depth $z_i$ in layer $j$, the simulated core level spectrum at $\theta_x$ is acquired with the dashed line representing $I_{j,\max}(E_b, \theta_x)$ in Equation (2), with $j$ = LCO. Finally, Fig. S8(e) is the full curve of the La 4d peak shift for all incidence angles, and this curve is in excellent agreement with experiment in Fig. 5(b). In arriving at the final linear potentials, slopes as shown in Fig. S8(b) have been varied, and the experimental core-level peak shifts with incidence angle compared to the calculations through a standard squared-intensity R factor to yield the optimum values. As a result of this simulation process, the slope of the built-in potential in each layer is determined individually.

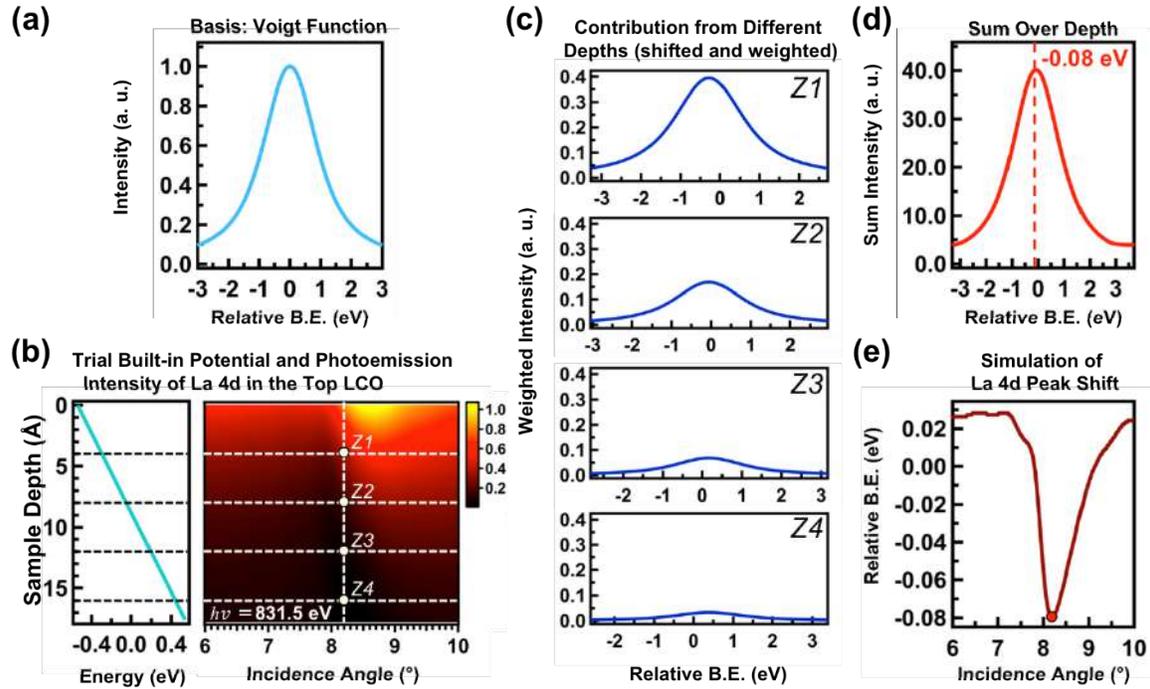

FIG. S8. Illustration of the method used to determine the linear segments of the built-in potential within each, for the example of La 4d in the LCO layer. (a) The Voigt function used as a basis function for simulating the peak shift of La 4d. Its Gaussian and Lorentzian line widths are set to be the same as the parameters extracted from the experimental curve fitting to the spectrum, and thus include instrumental and lifetime broadening. (b) Trial depth-resolved built-in potential and simulated photoemission intensity of the top LCO layer, with four reference depths Z1-Z4 shown. (c) Simulated peak contributions from the depths Z1-Z4 for x-ray incidence at the Bragg angle of 8.2°. The combined influence of the built-in potential and photoemission probing profile results in a peak shift and intensity variation with depth. (d) Simulated La 4d spectrum at the Bragg angle, calculated via summing over the contributions from all sample depths. Its peak position, at -0.08 eV with respect to a zero built-in potential, is marked by the red dashed line (e) Final simulated La 4d peak position versus angle. The red dot is the relative binding energy at the Bragg angle from panel (d).

**Simulation of the layer-projected valence spectra from bulk-reference XPS spectra—Figure S9**

In Fig. S9(a), the reference XPS spectrum from a thick sample of LCO is shown. We will treat this as representative of a bulk sample of LCO at our excitation energy. Although it should be noted that the relative differential cross sections of the different subshells contributing to these spectra could be somewhat different, if normalized to the cross section of Cr 3d for both energies, the Cr 3p/O 2p ratios are 3.86 at 825 eV and 3.85 at 1487 eV [5], and thus negligibly different. This reference spectrum we also assume not to vary with depth over the sensing depth of XPS and it is thus used as the "basis function" $I_{VB,j}^{XPS}(E_b - E_b^0(z_i))$ in the following equation for simulating the deconvoluted MEWDOSs in layer $j$ from the SW data:

$$I_{VB,j}(E_b) = \sum_{\theta_x} \sum_{z_i} I_{VB,j}^{XPS}(E_b - E_b^0(z_i)) |E(z_i,\theta_x)|^2 \exp(-z_i / \Lambda_e \sin\theta_e)$$

Figure S9(b) shows the built-in potential and the photoemission intensity estimate from $|E(z_i,\theta_x)|^2 \exp(-z_i / \Lambda_e \sin\theta_e)$, as in Fig. S8(b). Figure S9(c) is analogous to Fig. S8(c) in showing the individual depth-resolved, energy shifted, and intensity-weighted spectra. Finally, Fig. S9(d), shows the sum over multiple depths, the curve that is also shown in the bottom panel of Fig. 5(d) of the main text as the simulation.

To represent the depth resolved built-in potential completely, four variables are involved: two slopes and two steps, with the latter assumed for simplicity to be sharp. As part of this VB analysis, the potential step at each interface is varied, while the slopes are fixed at the values derived from the core-level shifts.

By minimizing the energy difference of VB edges between experimental and simulated results, with the experimental values being indicated by the dashed lines in Fig. 4(c) which neglect the influence of peak D that we assume is associated with Cr diffusing into STO, the band alignments between LCO and STO layers at two kinds of charged interfaces are determined.

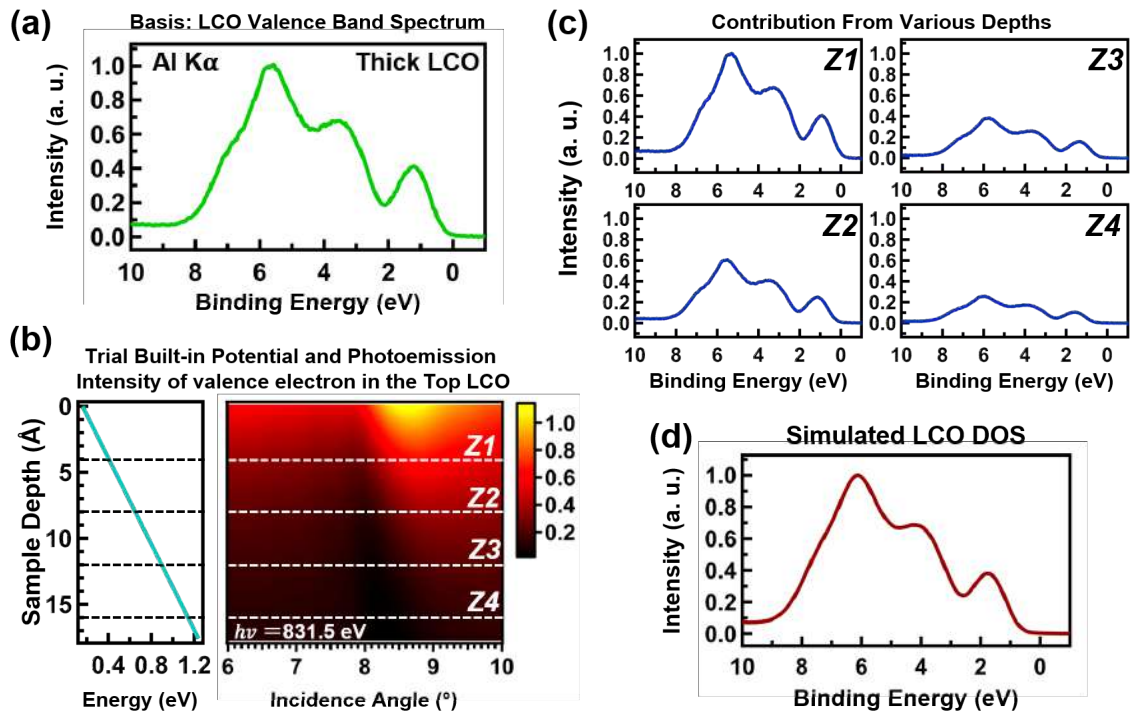

FIG. S9. As Figure S8, but for the example of the LCO valence spectrum. (a) Valence band spectrum of a thick LCO film acquired with Al Kα (1486.6 eV). This is used as the basis function of the simulation of the decomposed LCO MEWDOS, with a potential step at the interfaces included to properly align with the experimental VB maxima. (b) Trial depth-resolved

built-in potential and simulated photoemission intensity of the valence electrons in the top LCO layer. (c) Angle integrated contributions from various depths Z1 to Z4, as indicated by four dashed lines in (b). (d) Final simulated LCO MEWDOS.

**DFT modeling of the valence band maximum profile across the LCO/STO heterostructure- Figure S10:**

As discussed in a prior paper [6], the LCO/STO hetero-structure was represented using a periodic model with a $\sqrt{2}a_0 \times \sqrt{2}a_0$ lateral cell in the $a$–$b$ plane (where $a_0$ corresponds to the lattice parameter of a cubic perovskite lattice) and 15 unit cells (u.c.) along the c-axis: 5 u.c. of LCO followed by 10 u.c. of STO. The LaO/TiO$_2$ and SrO/CrO$_2$ interfaces are located at ~18 and ~58 Å in Figure S10. The internal coordinates and the lattice parameters for this system were optimized using the PBEsol density functional [7] as implemented in the Vienna Ab initio Simulation Package (VASP) [8,9]. The projector-augmented wave was used to approximate the electron-ion potential [10]; a 2×2×1 Monkhorst and Pack grid was used for Brillouin zone integration.

To investigate the dependence of the VB maximum profile on the details of the electronic structure, we performed simulations at the PBEsol+$U$ level to allow for correlation effects, where the Hubbard $U_{eff} = U - J$ correction was applied to Ti 3d and Cr 3d states [11]. Several combinations of $U_{eff}$ (Cr) and $U_{eff}$ (Ti) values were selected ($U_{eff}$ (Cr) = 0.0, 1.5, and 3.0 eV; $U_{eff}$ (Ti) = 0.0, 4.0, and 8.0 eV), where the maximum values of $U_{eff}$ provide close agreement between calculated and experimental band gaps for the LCO and STO bulk, respectively (see Figure S10). For consistency, the supercell parameters were fixed at the values derived using PBEsol (a = b =

5.48 Å; c = 58.34 Å), while the internal coordinates were re-optimized for each combination of $U_{eff}$.

The profile of the valence band maximum across the LCO/STO heterostructure was calculated using a procedure outlined elsewhere [12]. First, the electrostatic potential for the supercell was calculated on a three-dimensional grid and averaged in the a–b plane; a running average of the resulting potential along the c-axis was calculated and plotted as the lower blue curves in Figure S10. Then we calculated the shift of the VB maximum with respect to the average electrostatic potential in bulk STO and applied this shift throughout the entire STO region of the heterostructure; the same procedure was carried out for the LCO part of the heterostructure. Superposition of these contributions, merging half-way between the outer STO and LCO planes (shown in S10) is then the prediction of the VB maximum profile deduced from the experimental data, which agrees very well with our experimental determination, especially for the choices of $U_{eff}$ (Ti) = 4.0 eV and $U_{eff}$ (Cr) = 1.5 eV, as discussed in the text.

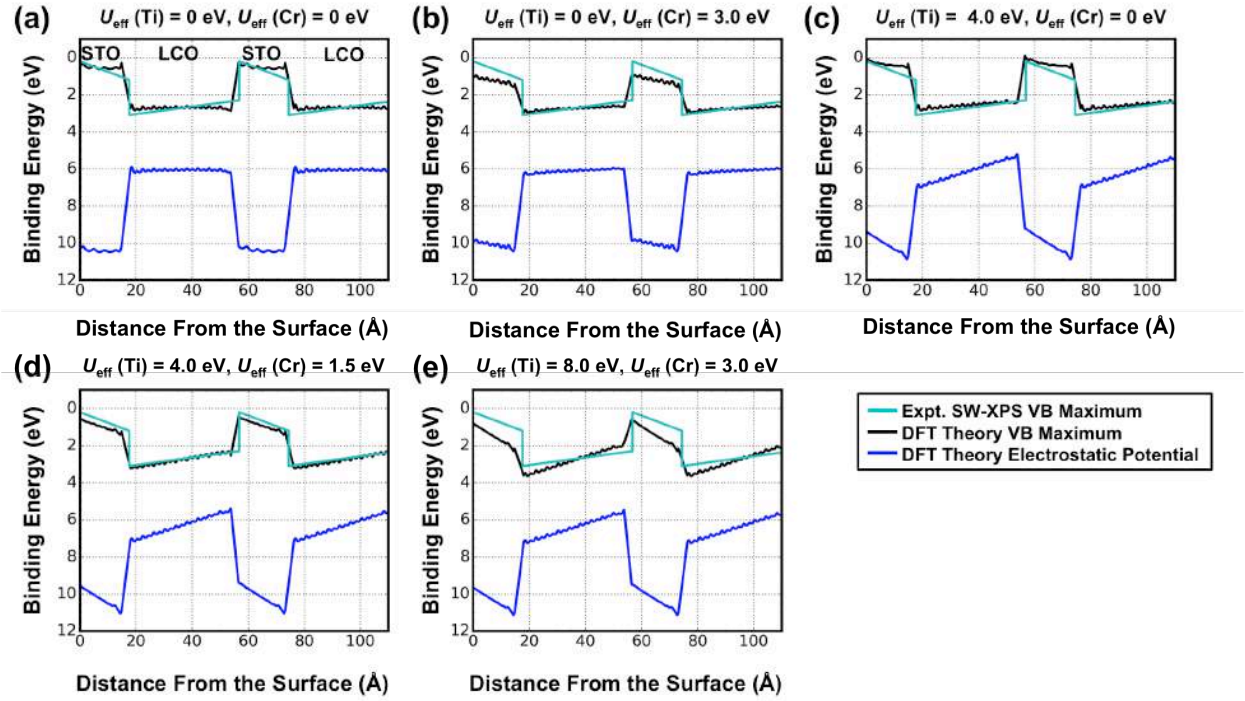

FIG. S10. DFT calculated (PBEsol+$U$ approach) depth-resolved electrostatic potentials of the LCO/STO superlattice with different sets of Hubbard U (blue curves) and their corresponding predicted valence band maxima (black curves). (a) $U_{eff}$ (Ti) = 0 eV, $U_{eff}$ (Cr) = 0 eV, (b) $U_{eff}$ (Ti) = 0 eV, $U_{eff}$ (Cr) = 3.0 eV, (c) $U_{eff}$ (Ti) = 4.0 eV, $U_{eff}$ (Cr) = 0 eV, (d) $U_{eff}$ (Ti) = 4.0 eV, $U_{eff}$ (Cr) = 1.5 eV and (e) $U_{eff}$ (Ti) = 8.0 eV, $U_{eff}$ (Cr) = 3.0 eV. The experimental SW-XPS derived VB maxima are plotted as the turquoise curves.

**Comparison between the SW-XPS derived band offsets and offsets calculated using the Kraut Method**

We also note that the VB offsets (VBOs) or steps in the potential in Fig. 5(e) are fully consistent with a much simpler global calculation neglecting multiple interfaces and inelastic attenuation, based on the method of Kraut *et al*.[13]. Here, we use the angle-averaged core-to-VBM differences from our data and similar differences from the XPS reference spectra of Chambers *et al*. [14] **Error! Bookmark not defined.**and calculate the band offset from the standard formula. Specifically, the VB simulations in Figs. 5(c) and 5(d) yield VBOs of 1.9 eV for the $LCO_{top}/STO_{bottom}$ interface and 2.1 eV at the $STO_{top}/LCO_{bottom}$ interface, whereas applying the Kraut method yields 2.0 eV for the above and below resonance cases. These numbers are in excellent agreement considering the different methods used and the fundamental difference in the samples: our multilayer with two distinct interfaces versus a simple bilayer with one interface of the $LCO_{top}/STO_{bottom}$ type in the case of Chambers *et al*.

**References:**

[1] A. X. Gray, C. Papp, B. Balke, S.-H. Yang, M. Huijben, E. Rotenberg, A. Bostwick, S. Ueda, Y. Yamashita, K. Kobayashi, et al., Interface properties of magnetic tunnel junction $La_{0.7}Sr_{0.3}MnO_3$/$SrTiO_3$ superlattices studied by standing-wave excited photoemission spectroscopy, Phys. Rev. B **82**, 205116 (2010).

[2] S. Nemšák, G. Conti, A. X. Gray, G. K. Palsson, C. Conlon, D. Eiteneer, A. Keqi, A. Rattanachata, A. Y. Saw, A. Bostwick, et al., Energetic, spatial, and momentum character of the electronic structure at a buried interface: The two-dimensional electron gas between two metal oxides, Phys. Rev. B **93**, 245103 (2016).

[3] Center for X-Ray Optics, Lawrence Berkeley National, Laboratory, http://henke.lbl.gov/optical_constants/getdb2.html.

[4] R. B. Comes, S. R. Spurgeon, D. M. Kepaptsoglou, M. H. Engelhard, D. E. Perea, T. C. Kaspar, Q. M. Ramasse, P. V. Sushko, and S. A. Chambers, Probing the origin of interfacial carriers in $SrTiO_3$–$LaCrO_3$ superlattices, Chem. Mater. **29**, 1147–1155 (2017).

[5] Non-relativistic photoelectric cross sections for all atoms, as calculated by J.J. Yeh and I. Lindau, J.J. Yeh and I.Lindau, Atomic Data and Nuclear Data Tables, 32, 1-155 (1985), and available at: https://vuo.elettra.eu/services/elements/WebElements.html.

[6] R. B. Comes, S. R. Spurgeon, S. M. Heald, D. M. Kepaptsoglou, L. Jones, P. V. Ong, M. E. Bowden, Q. M. Ramasse, P. V. Sushko, and S. A. Chambers, Interface-induced polarization in $SrTiO_3$-LaCrO3 superlattices, Adv. Mater. Interfaces **3**, 201500779 (2016).

[7] J. P. Perdew, A. Ruzsinszky, G. I. Csonka, O. A. Vydrov, G. E. Scuseria, L. A. Constantin, X. L. Zhou, and K. Burke, Restoring the density-gradient expansion for exchange in solids and surfaces, Phys. Rev. Lett. **100**, 136406 (2008).

[8] G. Kresse and J. Furthmüller, Efficient iterative schemes for ab initio total-energy calculations using a plane-wave basis set, Phys. Rev. B **54**, 11169 (1996).

[9] G. Kresse and J. Hafner, Ab initio molecular-dynamics simulation of the liquid-metal–amorphous-semiconductor transition in germanium, Phys. Rev. B **49**, 14251 (1994).

[10] P. E. Blöchl, Projector augmented-wave method, Phys. Rev. B **50**, 17953 (1994).

[11] S. L. Dudarev, G. A. Botton, S. Y. Savrasov, C. J. Humphreys, and A. P. Sutton, Electron-energy-loss spectra and the structural stability of nickel oxide: An LSDA+U study, Phys. Rev. B **57**, 1505 (1998).

[12] K. T. Delaney, N. A. Spaldin, and C. G. Van de Walle, Theoretical study of Schottky-barrier formation at epitaxial rare-earth-metal/semiconductor interfaces, Phys. Rev. B **81**, 165312 (2010).

[13] E. A. Kraut, R. W. Grant, J. R. Waldrop, and S. P. Kowalczyk, Precise determination of the valence-band edge in x-ray photoemission spectra: application to measurement of semiconductor interface potentials, Phys. Rev. Lett. **44**, 1620 (1980).

[14] S. A. Chambers, L. Qiao, T. C. Droubay, T. C. Kaspar, B. W. Arey, and P. V. Sushko, Band alignment, built-In potential, and the absence of conductivity at the $LaCrO_3/SrTiO_3$(001) heterojunction, Phys. Rev. Lett. **107**, 206802 (2011).